\documentclass[useAMS,usenatbib]{mn2e}

\usepackage{graphicx}
\usepackage{amssymb}
\usepackage{color}
\usepackage{natbib}

\title[Gamma-rays from electrons accelerated by MSPs in GCs]
{Numerical modelling of $\gamma$-ray emission produced by electrons originating from the magnetospheres 
of millisecond pulsars in globular clusters}
\author[A. Zajczyk, W. Bednarek, B. Rudak]{A. Zajczyk$^{12}$\thanks{E-mail:
azajczyk@physics.wustl.edu}, W. Bednarek$^{3}$, B. Rudak$^{1}$ \\
$^{1}$Physics Department, Washington University in St. Louis, 1 Brookings Drive, CB 1105, St. Louis, MO 63130, USA\\
$^{2}$Department of Astrophysics, N. Copernicus Astronomical Center, ul. Rabia\'{n}ska 8, 
87-100 Toru\'{n}, Poland\\
$^{3}$Department of Astrophysics, University of \L{}\'{o}d\'{z}, ul. Pomorska 149/153, 
90-236 \L{}\'{o}d\'{z}, Poland
}
\begin{document}

\date{Accepted 20... Received 2012}

\pagerange{\pageref{firstpage}--\pageref{lastpage}} \pubyear{2002}

\maketitle

\label{firstpage}

\begin{abstract}
Globular clusters are source of $\gamma$-ray radiation. At GeV energies, their emission is attributed
to magnetospheric activity of millisecond pulsars residing in the clusters. Inverse Compton scattering (ICS)
of ambient photon fields on relativistic particles diffusing through cluster environment is thought to be 
the source of GeV to TeV emission of globular clusters.
Using \emph{pair starved polar cap} model $\gamma$-ray emission from synthetic millisecond pulsar
was modelled. In addition to pulsar emission characteristics, the synthetic pulsar model yielded spectra
of electrons escaping pulsar magnetosphere.
To simulate $\gamma$-ray emission of globular cluster, both products of synthetic millisecond pulsar
modelling were used. Gamma-ray spectra of synthetic millisecond pulsars residing in the cluster were
summed to produce the magnetospheric component of cluster emission. Electrons ejected 
by these pulsars were injected into synthetic globular cluster environment. Their diffusion and interaction, both,
with cluster magnetic field and ambient photon fields, were performed with \citet{bednarek07} model yielding
ICS component of cluster emission. The sum of the magnetospheric and ICS components gives the synthetic
$\gamma$-ray spectrum of globular cluster. The synthetic cluster spectrum stretches from GeV to TeV energies.
Detailed modelling was preformed for two globular clusters: Terzan 5 and 47 Tucanae. Simulations are
able to reproduce (within errors) the shape and the flux level of the GeV part of the spectrum observed for both 
clusters with the Fermi/LAT instrument. The synthetic flux level obtained in the TeV part of the clusters'
spectrum is in agreement with a H.E.S.S. upper limit determined for 47 Tuc, and with emission level recently 
detected for Ter~5 with H.E.S.S. telescope. The synthetic globular cluster model, however, is not
able to reproduce the exact shape of the TeV spectrum observed for Ter 5.
\end{abstract}


\section{Introduction}
\label{gcs-intro}

Recently, globular clusters (GCs) were established as a source of GeV emission. So far,
the LAT instrument onboard the Fermi Space Telescope has detected 15 globular
clusters \citep{abdo09, abdo10, kong10, tam11} emitting photons in the energy range from $\sim 200$~MeV
to $\sim 50$~GeV. All of them are seen as point sources. 

For 8 of the $\gamma$-ray emitting GCs 
\citep{abdo10} the placement of the emission is coincident with the position of the clusters. 
In these cases the observed $\gamma$-ray emission does not vary in time. Moreover, their differential photon
spectra can be fitted with a power law with an exponential cut-off
The derived spectral indices are $\Gamma \lesssim 2.0$, and the cut-off energies $\mathcal{E}_{\rm c}$ 
fall in the range between 1 GeV and 4.5 GeV. However, for two of these GCs (M62 and NGC 6652) the determined 
spectral cut-offs are not statistically significant \citep{abdo10}. Nonetheless, the observed characteristics 
of the $\gamma$-ray emission of globular clusters resembles that observed for millisecond pulsars (MSPs)
detected with the Fermi/LAT instrument at GeV energies \citep{abdo09msp}. Thus, the population of MSPs residing 
in the clusters can be responsible for their $\gamma$-ray radiation.

\citet{tam11} reported detection of 7 more $\gamma$-ray emitting globular clusters. Interestingly, their 
$\gamma$-ray emission is slightly shifted with respect to the position of the cluster. The extreme case here is NGC~6441 
for which the maximum of the GeV emission falls far outside the cluster tidal radius \citep[see fig.~4 of][]{tam11}.
The $\gamma$-ray emission of these 7 GCs does not vary in time. However, in contrast with the clusters
detected by \citet{abdo10}, their observed photon spectra are best fitted by a power law. The exception
here is the spectrum of NGC~6441 that is better described by the exponentially cut-off power law
\citep{tam11}. For two of the clusters (Liller 1 and NGC 6624) the $\gamma$-ray spectra extend
to energies above 40 GeV. The power-law character of these spectra and their energy extent 
cannot be solely attributed to the magnetospheric emission of MSPs residing in these clusters. 
Possibly, the observed emission from these two GCs results from the inverse Compton scattering (ICS) of the ambient
photon fields on relativistic electrons injected into globular cluster environment by MSPs.

In addition to the Fermi/LAT detection in the GeV domain, the $\gamma$-ray emission at 
the TeV energies was detected towards Terzan~5 globular cluster \citep{gchess}. It is slightly 
extended beyond the GC tidal radius. Moreover, the peak of the emission is significantly shifted 
with respect to the cluster centre \citep[see fig.~1 of][]{gchess}. 
The observed spectrum can be approximated by a power law with a photon spectral index $\Gamma \simeq 2.5$. 
The ICS photons may substantially contribute to the observed 
TeV radiation from Terzan 5. However, the ICS origin of TeV emission cannot account for the observed
source morphology. \citet{gchess} give different possible explanations of the TeV radiation detected from
the direction of Ter 5. Source coincidence is one of the explored possibilities. In such case
the TeV $\gamma$-rays could originate from a pulsar wind nebula associated with a radio-quiet
pulsar. However, the probability of such scenario is very low. Other possibility
is that the TeV emission could have hadronic origin \citep[for details see][]{gchess}.
This scenario could explain the power-law emission spectrum. No TeV emission from the direction
of other globular clusters has been detected so far \citep{anderhub09, 47tuchess, mccut09}.

\citet{hui11} studied these 15 globular clusters detected in the $\gamma$-rays \citep{abdo09, 
abdo10, kong10, tam11} in search of correlations between the cluster $\gamma$-ray luminosity
$L_{\gamma}$ and various cluster properties. Such correlations could possibly facilitate explanation
of the origin of the $\gamma$-ray emission of these GCs. \citet{hui11} investigated the correlation
between $L_{\gamma}$ of the GC and two-body encounter rate $\Gamma_{\rm c}$, metallicity [Fe/H], absolute 
visual magnitude, and photon densities of the infrared and the optical Galactic background. The weakest 
correlation is found between $L_{\gamma}$ and the absolute visual magnitude of the cluster. However, strong
correlation is found between $L_{\gamma}$ and the two-body encounter rate and also the metallicity
of the cluster. The reported correlations are consistent with results previously found by \citet{hui10}
and \citet{abdo10}. Moreover, they find that the cluster $\gamma$-ray luminosity increases with 
the increase of the density of the background photon fields, both the optical and the infrared ones.
The correlation of $L_{\gamma}$ and $\Gamma_{\rm c}$ suggests that the $\gamma$-ray emission of
globular clusters is related to the population of objects formed through dynamical interactions
in the cluster. This in turn leads to the conclusion that $L_{\gamma}$ depends on the number
of MSPs residing in the cluster \citep{abdo10}. 

Metal-rich globular clusters are more likely to contain bright low-mass X-ray binaries (LMXBs). 
As proposed by \citet{ivanova06}, this can be related to the different stellar structure of main sequence 
donors with masses between $\sim 0.85 M_{\odot}$ and $\sim 1.25 M_{\odot}$. Metal-poor main-sequence stars do not
have an outer convective zone, while metal-rich ones do. Absence of the convective zone shuts
down magnetic braking, which in turn makes orbital shrinkage in these binaries less efficient. Thus,
preventing them from becoming bright X-ray sources, while opposite is the case for binaries with
metal-rich main-sequence donors. As low-mass X-ray binaries are thought to be progenitors of MSPs, 
higher probability of LMXBs formation for metal-rich globular clusters translates into a higher formation
rate of MSPs for these clusters. Thus, the observed correlation between $L_{\gamma}$ and [Fe/H] 
can link the origin of the $\gamma$-ray emission in GCs to the population of MSPs.

The observed correlation between $L_{\gamma}$ and the density of the photon fields cannot
be accounted for if the $\gamma$-ray radiation would originate solely from the magnetospheric
activity of MSPs. However, such correlation is expected when the $\gamma$-rays are produced
via ICS. Then the emitted power in ICS process is directly proportional to the energy 
density of the ambient photon field. The correlations found by \citet{hui11} \citep[and also][]{hui10, abdo10}
suggest that the interplay between the number of MSPs and the density of the background photon fields
determines the $\gamma$-ray luminosity of globular clusters.

\section{Theoretical approach to $\gamma$-ray emission of globular clusters}

In the past years, $\gamma$-ray emission from globular clusters was modelled assuming
it results from either combined magnetospheric radiation of millisecond pulsars residing
in a cluster, or from IC scattering of different photon fields on relativistic electrons
diffusing through GC interior.

The latter scenario was studied in detail by \citet[][hereafter BS07]{bednarek07}. In their model, BS07
assumed that relativistic electrons originate either in magnetospheres of pulsars populating
a globular cluster or are accelerated at the shock waves arising from collisions of pulsar
winds. In the first case BS07 assumed monoenergetic spectra for the electrons; in the second 
case they used power-law spectra. In both cases, after being injected into the cluster 
environment the relativistic electrons diffuse through the cluster interacting with its 
magnetic field. Moreover, on their way out electrons up-scatter photons a\-ri\-sing from stellar 
population in the cluster and photons of the cosmic microwave background (CMB).
The modelling of BS07 predicts $\gamma$-ray spectra of GCs spanning from GeV 
to TeV energies.

Similarly to BS07, \citet{cheng10} modelled $\gamma$-ray emission from globular 
clusters as arising from ICS processes taking place in the globular cluster environment. 
However, it was assumed that the relativistic electrons originate only from magnetospheres
of millise\-cond pulsars residing in the cluster. The injected electron spectra were 
monoenergetic. The relativistic leptons in the model of \citet{cheng10} up-scatter not only 
the stellar photons and the CMB, but also soft photons originating in the Galaxy (i.e. infrared 
background and stellar photons from the galactic disk). The resultant $\gamma$-ray emission 
spans GeV to sub-TeV energies. It is important to point out that the model of \citet{cheng10}
predicts rather high sub-TeV flux for globular clusters. Moreover, it also predicts that 
the $\gamma$-ray emission should extend beyond the cores of GCs (a radius $> 10$ pc).

\citet{venter09} assumed that both processes, ICS and magnetospheric $\gamma$-ray emission 
of millisecond pulsars, contribute to the overall $\gamma$-ray brightness of globular clusters.
Under assumption that an accelerating electric field in the magnetospheres of MSPs is unscreened
\citep{muslimov92, harding98} due to insufficient pair creation, $\gamma$-ray emission from 
MSPs was simulated. In their work, \citet{venter09} used population of cluster MSPs as a
sole source of relativistic electrons in the GC environment. The energy spectra of
electrons injected into the cluster were computed using the same model of the pulsar 
magnetosphere as the one used for simulating $\gamma$-ray emission of MSPs. Using 
Monte Carlo method, average cumulative magnetospheric spectrum and cumulative electron 
injection spectrum originating from 100 MSPs were calculated. This allowed for calculating
GeV-to-TeV spectra for globular clusters 47 Tuc and Ter 5. Similarly to BS07,
to produce ICS component \citet{venter09} up-scattered the stellar photons and the CMB 
photons. 

We note that while modelling of GeV-TeV spectral characteristics of GCs puts
constraints on the possible number of MSPs populating the cluster and on the cluster 
magnetic field \citep[given energy densities of ambient photon fields and diffusion coefficients; e.g.][]{venter09},
X-ray energy domain yields additional constraints on physical properties of GC. 
In particular, \citet{buesching12} modelled synchrotron emission maps at radio
and X-ray energies for Terzan 5. Comparison of the modelled X-ray emission map 
with the detected diffuse X-ray emission from Ter 5 \citep{eger10} allowed to
put additional constraints on diffusion process of relativistic particles within GC.
For more details on the content of GCs, their high energy observations and modelling 
see recent review by \citet{bednarek11}.

In this paper we present the results of the numerical modelling of $\gamma$-ray radiation from globular
clusters. In our study it is assumed (similarly as in \citet{venter09}) that the cluster $\gamma$-ray emission results from 
combined magnetospheric activity of millisecond pulsars residing in the cluster, and from ICS
scattering of ambient photon fields on relativistic electrons propagating through the cluster.
These electrons are injected into the cluster environment by the MSPs. Detailed description of 
the synthetic millisecond pulsar model used for computing $\gamma$-ray spectra of pulsars residing 
in the cluster and also spectra of electrons ejected from their magnetospheres can be found in 
Sect.~\ref{msp-model}. In Sect.~\ref{synth-gc} details of calculations of synthetic $\gamma$-ray 
spectra of globular clusters are presented. Comparison of the simulations with the existing 
observational data for selected globular clusters can be found in Sect.~\ref{comp-sim-obs}. 
Conclusions are given in Sect.~\ref{conlude}.

\section{Numerical model of synthetic millisecond pulsar}
\label{msp-model}

In this section we discuss models of $\gamma$-ray production in the open magnetosphere 
(distances comparable to the light cylinder radius) of pulsars, and motivate why the \emph{pair 
starved polar cap} model \citep[PSPC;][]{muslimov04} is preferred by us for the MSPs residing 
in globular clusters. We also discuss results of PSPC model that are used to calculate magnetospheric 
and ICS components to the $\gamma$-ray GeV-TeV emission of synthetic globular cluster (Sect.~\ref{synth-gc}).

\subsection{Justification for the selected synthetic millisecond pulsar model}
In the standard polar cap model approach \citep[see e.g.,][]{harding98} the accelerating
electric field is being screened out at distances close to the neutron star surface ($\sim$ few stellar
radii), which means that the acceleration gap fills only a small fraction of the open magnetosphere.
Such description may be applicable to classical pulsars (with ages $\tau_{\mathrm{c}} \lesssim 10^{7}$~yr)
where in their magnetospheres pair production through interaction of curvature and/or inverse Compton
scattered photons with the magnetic field is sufficient to completely screen the electric field at low
altitudes \citep[see e.g.,][]{muslimov04}. However, the situation is completely different for much older
pulsars, including millisecond ones. For these pulsars the primary electrons may keep accelerating and
at the same time keep emitting high energy photons up to very high altitudes without significant pair
production \citep[see e.g.,][]{muslimov04, harding05}. \citet{harding05} show that for millisecond
pulsars the primaries can be accelerated up to distances comparable to the light cylinder radius.

The characteristics of some of the $\gamma$-ray light curves of the millisecond pulsars and their behaviour 
with respect to the radio light curves \citep[i.e. misalignment of $\gamma$-ray peak and radio peak;][]{abdo09msp} 
seem to favour outer magnetosphere as a region where the $\gamma$-rays originate. However, the $\gamma$-ray 
and radio light curve modelling performed by \citet{venter09} and \citet{venter12} showed that the group 
of $\gamma$-ray bright MSPs is non-uniform in terms of the preferred emission model. 
Three subclasses can be distinguished - two of the subclasses favour the outer magnetosphere as
the place of origin of the $\gamma$-ray radiation (either in the framework of the outer gap or 
the slot gap model). However, there are MSPs whose $\gamma$-ray emission can only be explained 
invoking the extended polar cap model, i.e. the \emph{pair starved polar cap} model \citep{muslimov04} 
where the $\gamma$-ray emission is produced in the whole volume of the pulsar open magnetosphere.

In the slot gap model \citep{muslimov04slg} the approximate solution to the accelerating electric field $E_{||}$
suffers from a sign reversal. In the outer gap model non-dipolar magnetic field \citep{cheng00} needs 
to be invoked in order to obtain copious electron-positron pair creation. In turn, the created pairs are needed 
to control the gap dimensions. The pair starved polar cap model, similarly to the slot gap model, suffers from 
the sign reversal in the solution for the unscreened accelerating electric field $E_{||}$ (see Sect. 3.3). 
Nonetheless in the past years it was widely studied \citep[see e.g.,][and references therein]{venter09} 
in the context of $\gamma$-ray emission of millisecond pulsars.

Because properties of MSPs in globular clusters are different from the ones of the field population \citep{camilo05} 
it is justified to suspect that the $\gamma$-ray emission mechanism operating in the magnetospheres of these 
globular cluster MSPs may be different from the one operating in the classical pulsars and/or in the field 
millisecond pulsars. Thus, in the presented study the \emph{pair starved polar cap} (PSPC) model is chosen to simulate
the $\gamma$-ray emission of millisecond pulsars.

\subsection{Basic assumptions of the synthetic millisecond pulsar model}
To simulate the $\gamma$-ray emission of millisecond pulsars a 3 dimensional (3D) numerical model
of pulsar magnetosphere was used \citep{dyksphd}. The model operates within a \emph{space charge
limited flow} framework. The magnetic field of a pulsar is in the form of a retarded vacuum dipole
\citep{dyks04} for which the curvature radii of magnetic field lines are determined in the inertial 
frame of reference. Moreover, following \citet{dyks02} the special relativity effects like aberration 
and time-of-flight delays are treated accordingly throughout the calculations.

Important change introduced into the numerical code was an inclusion of a more realistic 
description of the electric field $E_{||}$ (for details see Sect.~\ref{eacc}).
It is assumed that acceleration of particles takes place in the whole 
volume determined by the last open magnetic field lines. The electrons are injected at the polar cap 
surface at the Goldreich-Julian rate, and they have low initial energy. As there is no screening
of the electric field, like in the standard polar cap model \citep[see e.g.,][]{harding98}, acceleration
of the particles takes place even at distances comparable to the light cylinder radius. Such model
is called a \emph{pair starved polar cap} model \citep{muslimov04}. The implemented electric
field takes into account the general relativistic effect of dragging of inertial frames \citep{muslimov92,
muslimov04}.

\subsection{Structure of the accelerating electric field}
\label{eacc}
 
In the 3D numerical model of the millisecond pulsar magnetosphere the accelerating electric field $E_{||}$ in 
the form proposed by \citet{muslimov04} was implemented. The field itself extends up to the light cylinder 
and consists of three different prescriptions ($E_{1}$, $E_{2}$ and $E_{3}$) determining $E_{||}$ at different 
heights above the pulsar surface.

In the region close to the neutron star surface (further referred to as the \emph{near} regime) 
the accelerating electric field is described by the formula derived by \citet{muslimov92} under assumption that 
the electric field must disappear at the stellar surface ($E_{||} = 0$):
\begin{eqnarray}
	E_{1} & = & -3 E_{0} \theta_{0}^{2} ~\lbrace \kappa \cos \alpha \sum_{i=1}^{\infty} 
	\frac{4 J_{0}(k_{i}\xi)}{k_{i}^{3}J_{1}(k_{i})} [1-e^{-\gamma_{i}z}] \nonumber \\
	& + & \theta_{0} H(1) \delta(1) \sin \alpha \cos \phi \sum_{i=1}^{\infty} 
	\frac{2 J_{1}(\tilde{k}_{i} \xi)}{\tilde{k}_{i}^{3} J_{2}(\tilde{k}_{i})} [1-e^{-\tilde{\gamma}_{i} z}]
	\rbrace .
	\label{enear}
\end{eqnarray}
The above equation is expressed in the \emph{magnetic} coordinate system ($r$, $\theta$, $\phi$),
where $r$ is the radial distance from the neutron star centre, $\theta$ is the magnetic colatitude, and $\phi$
is the magnetic azimuthal coordinate. This is also the case for all the $E_{||}$ formulae presented in this Section. 
The angle $\alpha$ in Eq.~\ref{enear} is the angle between the rotation and the magnetic axis, $\xi \equiv 
\theta/\theta(\eta)$ is the magnetic colatitude scaled with the polar angle of the last open magnetic 
field line, $\eta = r/R_{\mathrm{NS}}$ is the radial distance scaled with the star radius, and $z \equiv \eta -1$ 
is the altitude above the neutron star surface scaled with the star radius.
The electric field is derived using a small-angle approximation, in which the polar angle of the last open 
magnetic field line $\theta(\eta) \ll 1$. This results in:
\begin{equation}
	\theta(\eta) \simeq \theta_{0} \sqrt{ \eta \frac{f(1)}{f(\eta)}} ~, ~~~\mathrm{where}~~~~
	\theta_{0} = \sqrt{\frac{\Omega R_{\mathrm{NS}}}{c f(1)}} ~. 
\end{equation}
Here $\theta_{0}$ is the magnetic colatitude (polar angle) of the base of the line at the stellar surface.
Other quantities used in Eq.~\ref{enear} are:
\begin{equation}
	E_{0} \equiv B_{\mathrm{s}}\frac{\Omega R_{\mathrm{NS}}}{c} ~, ~~~~~ 
	\kappa \equiv \frac{\epsilon I}{M_{\mathrm{NS}} R_{\mathrm{NS}}^{2}} ~, ~~~~~
	\epsilon \equiv \frac{2 G M_{\mathrm{NS}}}{c^{2} R_{\mathrm{NS}}} ~,
\end{equation}
where $M_{\mathrm{NS}}$ is the mass of a neutron star, $\epsilon = r_{g}/R_{\mathrm{NS}}$ is the
compactness parameter ($r_{g} = 2GM_{\mathrm{NS}}/c^{2}$ is the gravitational radius of a neutron star),
and $\kappa$ is the magnitude of the general relativistic effect of the frame dragging at the stellar surface
measured in the stellar angular velocity $\Omega$. The function $J_{m}$ is a Bessel function of order $m$. The
$k_{i}$ and $\tilde{k}_{i}$ are the positive roots of the Bessel functions $J_{0}$ and $J_{1}$. Functions
$\gamma_{i}$ and $\tilde{\gamma}_{i}$ are defined as:
\begin{equation}
	\gamma_{i} \approx \frac{k_{i}}{\theta_{0}(1-\epsilon)^{1/2}}~, ~~~\mathrm{and}~~~~
	\tilde{\gamma}_{i} \approx \frac{\tilde{k}_{i}}{\theta_{0}(1-\epsilon)^{1/2}}~.	
\end{equation}
Formulae for $f(\eta)$, $H(\eta)$ and $\delta(\eta)$ were first defined in \citet{muslimov92} and can
be found therein. The \emph{near} regime solution $E_{1}$ is applicable for altitudes $\eta -1 \ll 1$.

At moderate heights ($\theta_{0} \ll \eta -1 \ll \frac{c}{\Omega R_{\rm NS}}$; further referred to as 
the \emph{moderate} regime) the electric field is given by eq.~14 of \citet{harding98}:
\begin{eqnarray}
	E_{2} & = & -\frac{3}{2} E_{0}\theta_{0}^2 [ \frac{\kappa}{\eta^{4}} \cos \alpha \nonumber \\
		& + & \frac{1}{4} \theta(\eta) H(\eta) \delta(\eta) \xi \sin \alpha \cos \phi ]
		(1 - \xi^{2}) ~.
	\label{emoder}
\end{eqnarray}

Close to the light cylinder (further referred to as the \emph{far} regime) the electric field 
is described by eq.~35 of \citet{muslimov04}:
\begin{eqnarray}
	E_{3} & \approx & -\frac{3}{16} \left( \frac{\Omega R_{\mathrm{NS}}}{c} \right)^{3} \frac{B_{\mathrm{s}}}{f(1)}
		~[ \kappa \left(1-\frac{1}{\eta_{c}^{3}}\right) (1 + \xi^{2}) \cos \alpha \nonumber \\
		& + & \frac{1}{2} (\sqrt{\eta_{c}} - 1) \left( \frac{\Omega R_{\mathrm{NS}}}{c} \right)^{1/2} \lambda (1+2\xi^{2})
		\xi \sin \alpha \cos \phi ] \nonumber \\
		& \times & (1-\xi^{2}) ~,
	\label{efar}
\end{eqnarray}
where $\lambda = H(1)/\sqrt{f(1)}$, and $\eta_{\mathrm{c}}$ is a radial parameter at which the electric field 
saturates \citep{muslimov04slg}.

In order to obtain a smooth transition between the electric field working in the \emph{moderate} regime
(Eq.~\ref{emoder}) and the one working in the \emph{far} regime (Eq.~\ref{efar}) the formula proposed by 
\citet{muslimov04} is used:
\begin{equation}
	E_{||} \simeq E_{2} \exp\left[-\frac{\eta-1}{\eta_{\mathrm{c}}-1} \right] + E_{3} ~.
\label{eqtot}
\end{equation}
The value of $\eta_{\rm c}$ parameter is determined for each magnetic field line separately. \citet{muslimov04} 
showed that for millisecond pulsars the best-fit value of $\eta_{c}$ falls in the range $\sim 3 - 4$. Moreover,
they point out that for a given pulsar spin period the parameter $\eta_{c}$ is a function of an inclination
$\alpha$ and an azimuthal coordinate $\phi$ at the polar cap. 

In the numerical code before electrons are injected into the pulsar magnetosphere, we find the accelerating
electric field solution in the whole volume of the open magnetosphere up to the light cylinder distances.
When finding the full $E_{||}$ solution that could accelerate particles also in the distant parts of the
magnetosphere (so a combination of Equations~\ref{enear}, \ref{emoder} and \ref{efar} with the use of 
Eq.~\ref{eqtot}), following \citet{venter09}, we require that the resultant electric field:
\begin{description}
	\item - should be negative for all $\eta$ in the range between 1 and $\eta_{\mathrm{LC}}$, 
	where $\eta_{\mathrm{LC}} = r/R_{\mathrm{LC}}$;
	
	\item - should transit from the \emph{near} to the \emph{moderate} $E_{||}$ regime 
	(matching $E_{1}$ with $E_{2}$) when $E_{1} \simeq E_{2}$;
	
	\item - should transit from the \emph{moderate} to the \emph{far} $E_{||}$ regime 
	(matching $E_{2}$ with $E_{3}$) smoothly, which is achieved by applying Eq.~\ref{eqtot} and appropriate 
	matching of $\eta_{c}$;
	
	\item - should converge to $E_{3}$ for the distances close to the light cylinder.	
\end{description}
For the implemented electric field (Eq.~\ref{enear}, \ref{emoder} and \ref{efar}) a sign reversal of 
the electric field occurs at some altitude. This happens especially for those magnetic field lines that
cross the null-charge surface\footnote{\emph{null-charge surface} is the surface in the pulsar magnetosphere 
separating volumes filled with the particles of opposite charge.} within the volume of the open magnetosphere 
limited by the light cylinder radius. When $E_{||}$ changes sign, the particle oscillations 
occur. In order to alleviate this somewhat problematic situation it is required that $E_{||}$ field is negative 
for all $\eta$. If this condition is not fulfilled for a certain magnetic field line, the line is left out from 
the calculations so no acceleration of particles takes place along this line. To obtain smooth transition 
from $E_{2}$ to $E_{3}$ we take a broader range of possible values of $\eta_{c}$ than the one found by 
\citet{muslimov04} for millisecond pulsars. The values of $\eta_{c}$ are chosen from the range between 1 and 6. 
In the matching procedure the largest possible value of $\eta_{c}$ that ensures the smooth transition 
between the electric field solutions given by Eqs.~\ref{emoder} and \ref{efar} is chosen.

\begin{figure}
	\begin{center}
	\begin{tabular}{cc}
	\hspace{-0.45cm}\includegraphics[scale=0.3]{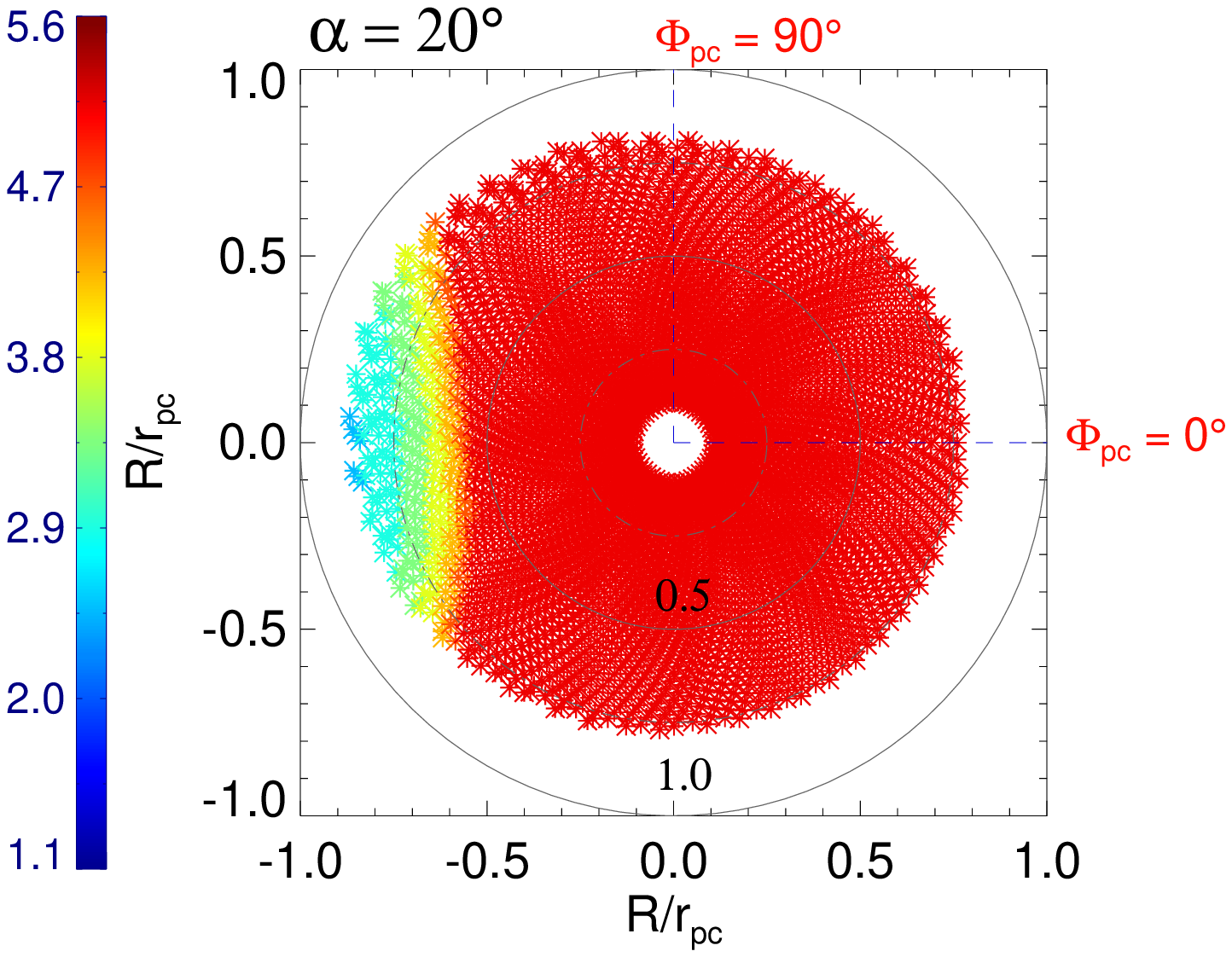}&
	\hspace{-1.3cm}\includegraphics[scale=0.3]{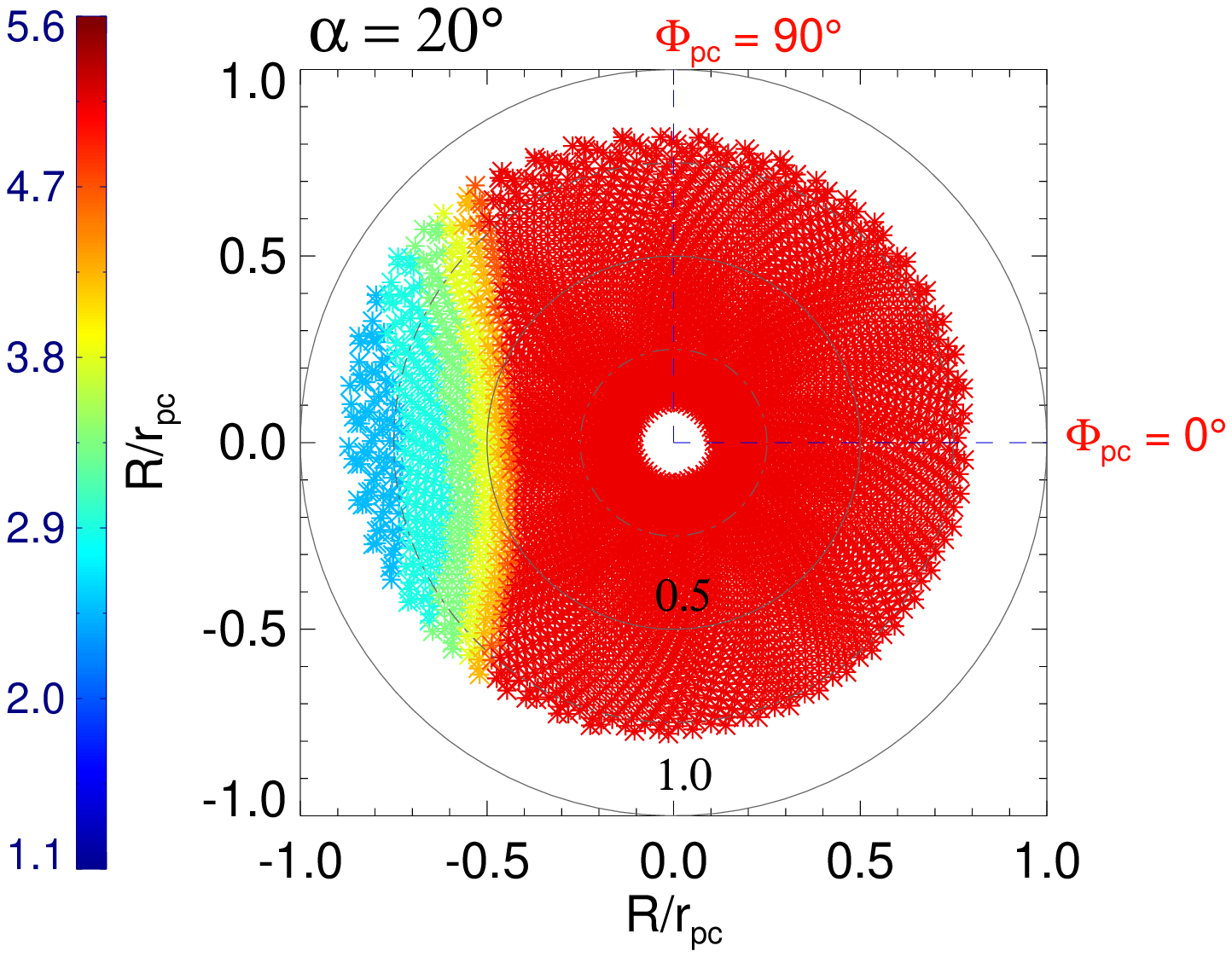}\\
	\hspace{-0.45cm}\includegraphics[scale=0.3]{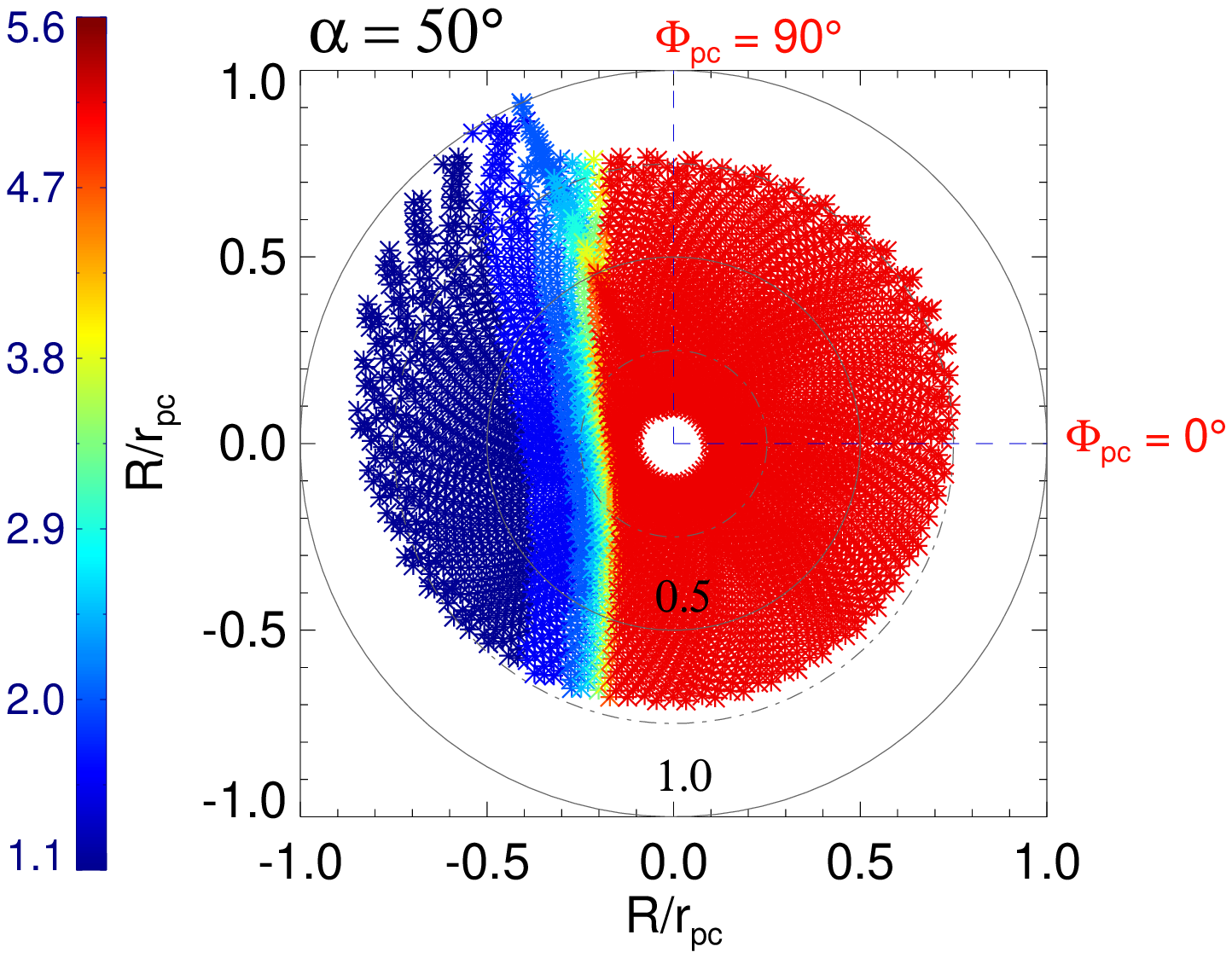}&
	\hspace{-1.3cm}\includegraphics[scale=0.3]{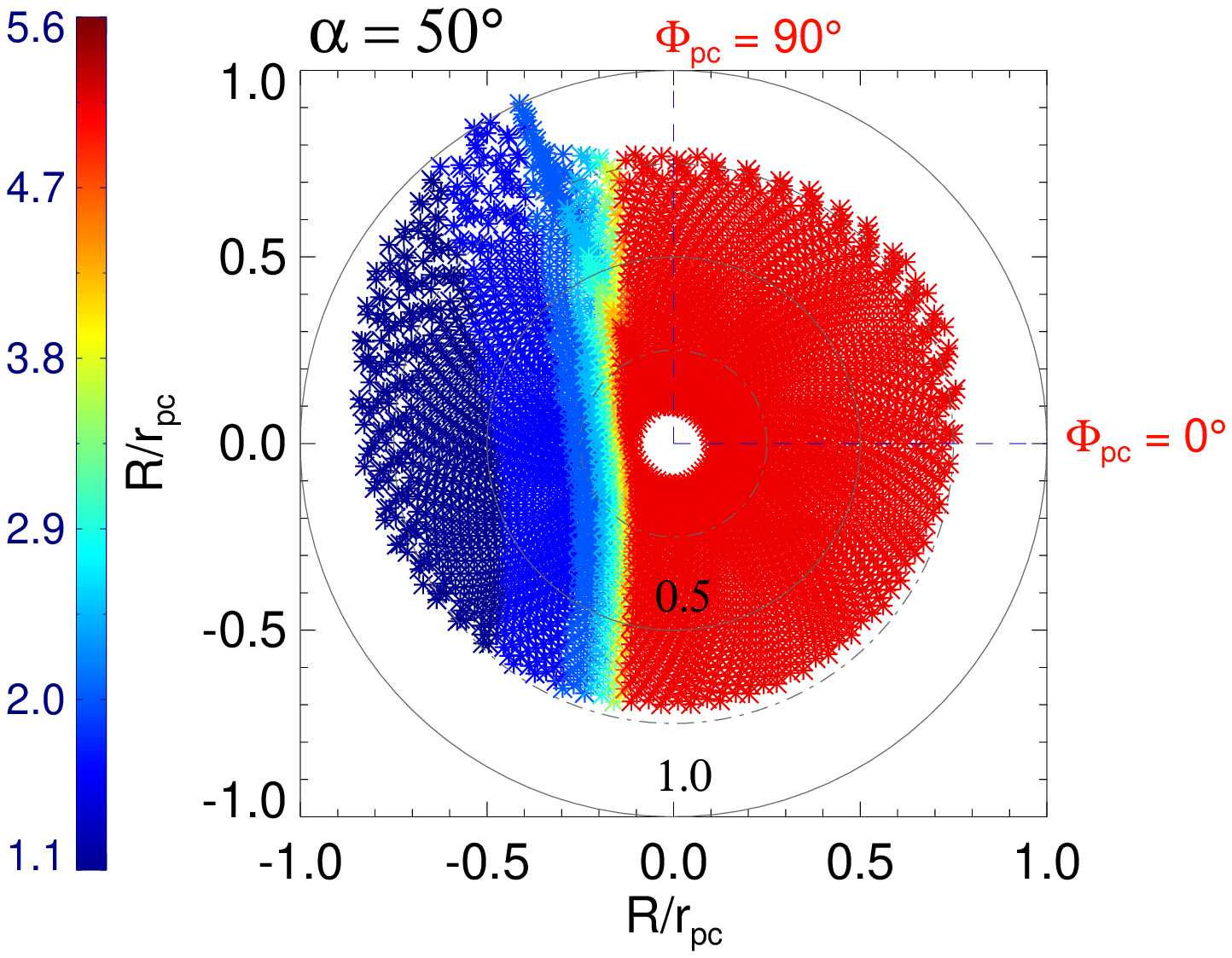}\\
	\hspace{-0.45cm}\includegraphics[scale=0.3]{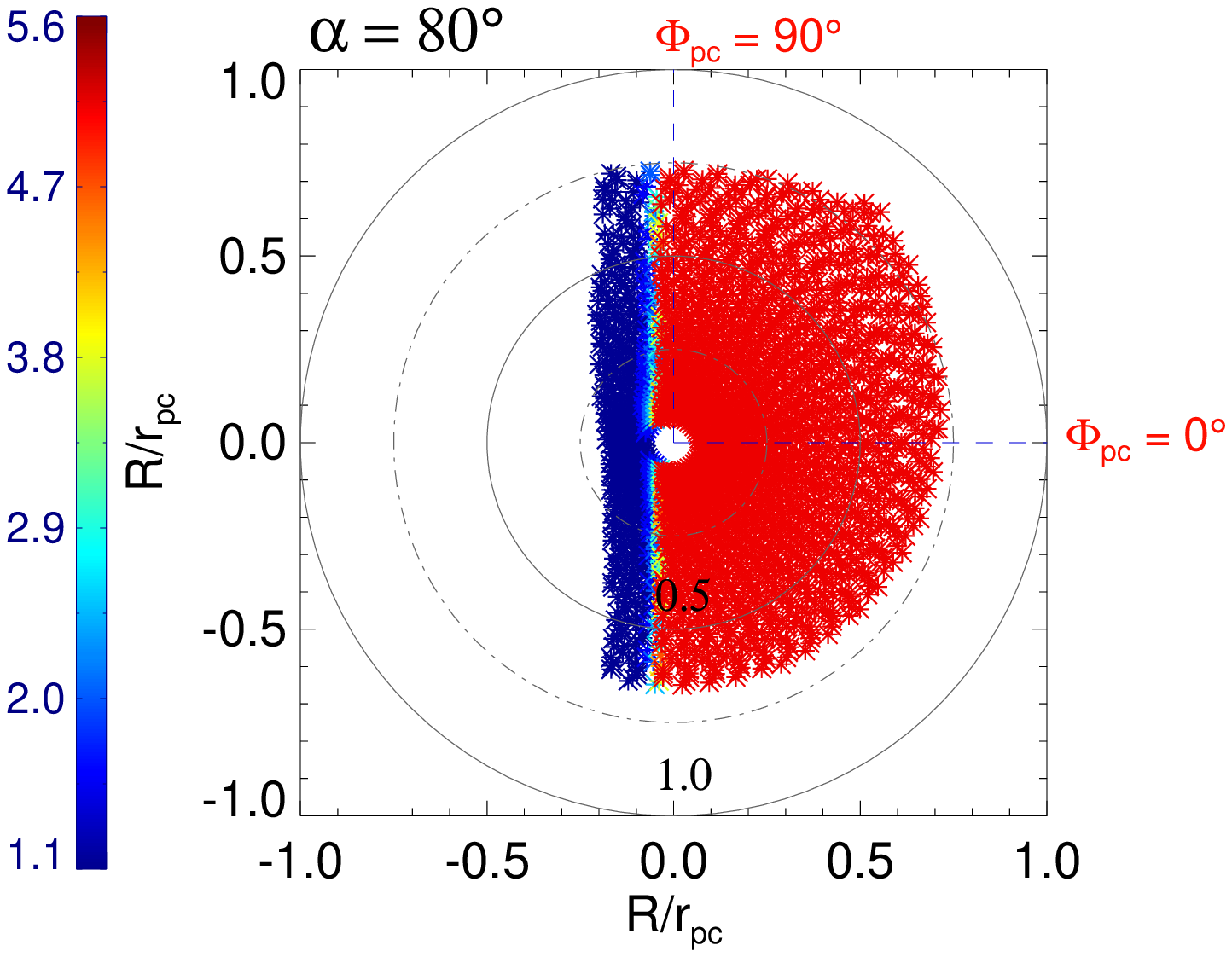}&
	\hspace{-1.3cm}\includegraphics[scale=0.3]{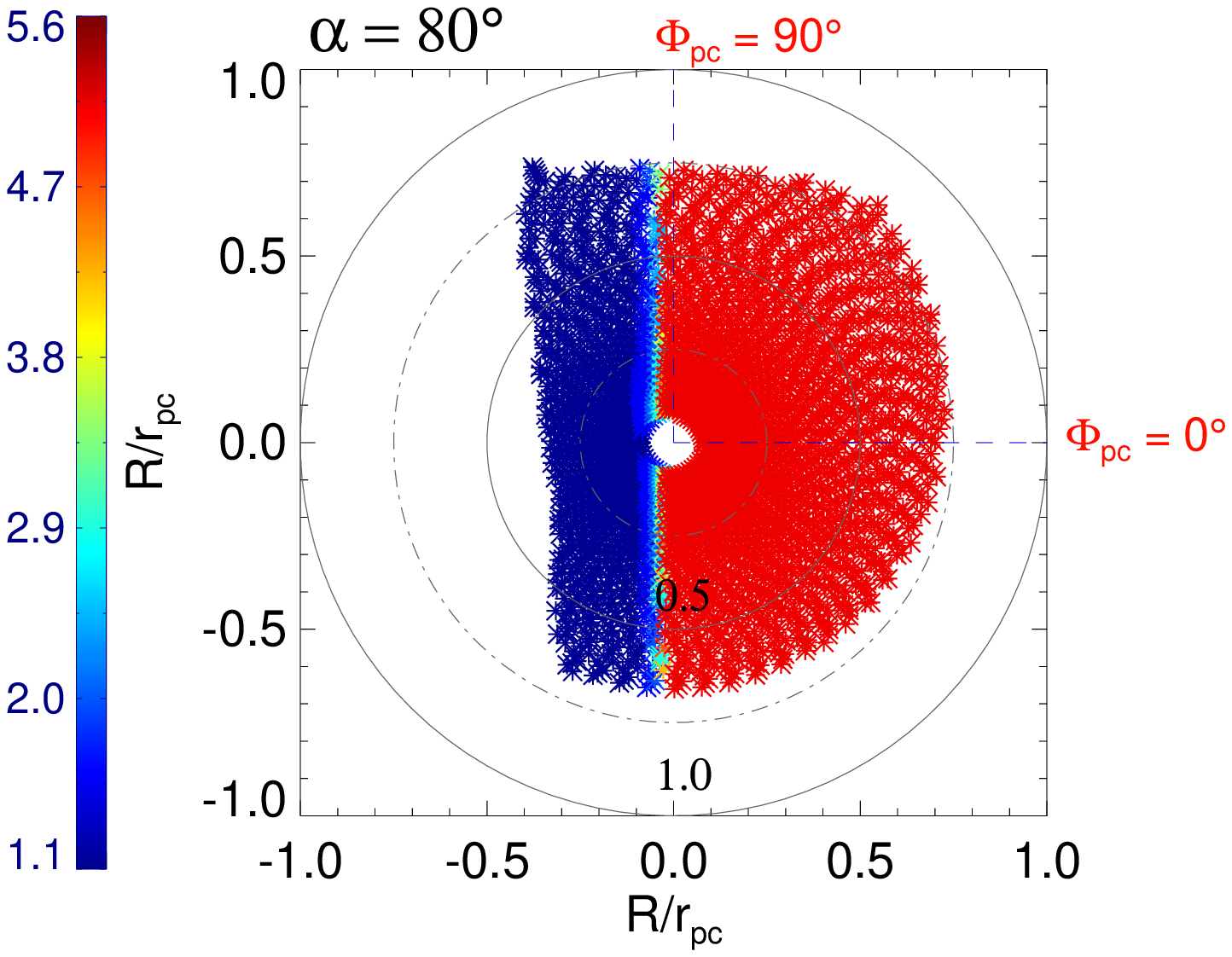}\\
	\end{tabular}
	
	\caption{\small The distribution of $\eta_{c}$ parameter across the pulsar polar cap.
	The images show results for synthetic MSP with $B_{\mathrm{s}} = 10^{8}$~G. 
	The \emph{left} column shows the case with $P = 1.5$~ms, while the \emph{right} column shows the
	case with $P = 9.5$~ms. The inclination changes along columns: \emph{top} row $\alpha = 20^{\circ}$, 
	\emph{middle} row $\alpha = 50^{\circ}$, and \emph{bottom} row $\alpha = 80^{\circ}$. Different values 
	of the parameter $\eta_{c}$ are colour-coded (see colorbars for the exact values).
	White regions are those for which no $E_{||}$ solution is found. 
	Small and big solid circle mark the distance across the polar cap equal to $0.5r_{\mathrm{pc}}$ 
	and $1.0r_{\mathrm{pc}}$, respectively. Here $r_{\mathrm{pc}}$ is the polar cap radius. Horizontal 
	red dashed line shows $\phi=0^{\circ}$ while vertical one points to $\phi=90^{\circ}$ ($\phi$ is 
	the magnetic azimuth angle).} 
	\label{efield-etac-dist1}
	\end{center}
\end{figure}

The distribution of the $\eta_{c}$ parameter across the polar cap for the case of $B_{\mathrm{s}} = 10^{8}$
and different values of $P$ and $\alpha$ is presented in Fig.~\ref{efield-etac-dist1}. When the inclination 
increases a region with progressively smaller values of $\eta_{c}$ increases at the polar cap side facing away 
from the rotation axis ($\phi = 180^{\circ}$). This is true for all presented spin periods.
For $\alpha = 80^{\circ}$ large part of the magnetosphere is excluded from the calculations (no solution
of $E_{||}$ is found). Such behaviour results from the fact that when we increase pulsar inclination 
the magnetic field lines situated in the part of the magnetosphere at large distances from the rotation
axis start crossing the null-charge surface. Thus, in order to mitigate the problem of particle oscillations
the open magnetic field lines crossing the null-charge surface are excluded from the calculations.

Previously, the $\gamma$-ray emission of millisecond pulsars was modelled by e.g. \citet{venter05} and
\citet{zajczyk08}. Their calculations were carried out in the framework of the extended polar cap model.
However, they considered accelerating electric field $E_{||}$ of the form given by Eqs.~\ref{enear} 
and \ref{emoder}. Thus, electrons were accelerated only up to the distances $\ll \eta_{\rm LC}$. 
\citet{venter09} performed the calculations of the $\gamma$-ray characteristics of the millisecond pulsars using the 
pair starved polar cap model, thus including also Eq.~\ref{efar} in the description of the accelerating 
electric field.
The conditions that need to be met when finding the solution of $E_{||}$ along the magnetic field 
lines are similar for \citet{venter09} and the implementation of the PSPC model presented in this work.
However, one difference
can be found. In order to match the electric field solutions in the \emph{near - moderate} distance regime 
\citet{venter09} used the approximate formula $\eta_{\rm b} \approx 1 + 0.0123 P^{-0.333}$, where 
$\eta_{\rm b}$ is the distance at which the transition between $E_{1}$ and $E_{2}$ takes place.
In this case, the distance $\eta_{\rm b}$ is independent of the position at the polar cap, so the
change from $E_{1}$ to $E_{2}$ for all open magnetic filed lines takes place at the same distance
from the neutron star surface. In the presented study, however, the distance of transition between 
$E_{1}$ and $E_{2}$ is determined for each of the magnetic field lines independently. The transition
takes place at the distance where the condition of $E_{1} \simeq E_{2}$ is fulfilled. Thus, the 
transition height from the \emph{near} to \emph{moderate} regime varies across the polar cap.
This in turn results in different transition heights from the \emph{moderate} to \emph{far} regime,
so the slightly different $\eta_{\rm c}$ values, with respect to the ones found by \citet{venter09}. 
The general behaviour of the $\eta_{\rm c}$ parameter with the increasing inclination is similar in both cases.
Additionally, the step in the numerical grid of 
the $\eta_{\rm c}$ values used in the $E_{||}$ calculations has direct impact on the exact distribution 
of $\eta_{\rm c}$ across the polar cap. Thus, the final result is the outcome of the two effects.
The observed differences between \citet{venter09} and the presented implementation of the PSPC model,
though observed, have minor effect on the calculated $\gamma$-ray emission characteristic of the synthetic
millisecond pulsars.

Having found the accelerating electric field structure across the pulsar magnetosphere, electrons are injected 
at the polar cap with low initial energy ($\gamma_{0} = 10$). The particles are distributed across the polar
cap ($\eta = 1$) with the Goldreich-Julian charge density that includes general relativistic corrections 
\citep[see e.g.,][]{muslimov97,harding98}:
\begin{eqnarray}
	\rho_{\mathrm{GJ}} & = & - \frac{\Omega B_{\mathrm{s}}}{2 \pi c \tilde{\alpha} \eta^{3}} \frac{f(\eta)}{f(1)} 
	[ \left(1 - \frac{\kappa}{\eta^{3}} \right) \cos \alpha \nonumber \\	
	& + & \frac{3}{2} \theta(\eta) H(\eta) \xi \sin \alpha 
	\cos \phi ] ~,
	\label{gj-gr}
\end{eqnarray}
where $\tilde{\alpha} \equiv \sqrt{1 - \epsilon/\eta}$ is the red-shift function. Their starting positions 
coincide with the foot-points of the magnetic field lines. Then the primary particles are being followed 
in their motion along the magnetic field lines. Moving through the pulsar magnetosphere the particles 
loose their energy through curvature radiation being at the same time accelerated by the electric field.

The emitted curvature photons interact with magnetic field. However, secondary pairs (electrons and positrons) 
are produced in insufficient numbers to screen out the electric field. The majority of curvature photons escape 
pulsar magnetosphere without being absorbed. At the same time, information on their energy and direction of motion 
is collected to produce photon maps \citep{dyksphd}, which give information on pulsar emission characteristics 
(the number of photons $d\dot{N}_{\gamma}$ radiated per unit time, per unit energy $d\mathcal{E}$, into unit 
solid angle $d\Omega$). The photon maps are used to 
obtain pulsar spectra and light curves for different observing angles $\zeta$.

Numerical calculations were performed for discrete distribution of pulsar parameters ($P$, $B_{\rm s}$, $\alpha$).
Pulsar spin period was selected from the set of $P = (1.5, 2.0, 3.5, 5.0, 6.5, 8.0, 9.5)$ ms. Magnetic field values 
were chosen from the set of $B_{\rm s} = (1.0, 3.5, 4.5, 5.5, 7.5, 9.0) \times 10^{8}$ G. Pulsar inclination angle was 
chosen from a range between $10^{\circ}$ and $80^{\circ}$ with a step of $10^{\circ}$. Such parameter selection
allowed for obtaining 336 different synthetic millisecond pulsar cases. For each modelled case a $\gamma$-ray
photon map\footnote{\emph{photon map} yields information on pulsar emission characteristics (the number of 
photons radiated per unit time, per unit energy, into unit solid angle) as a function of pulsar rotation phase $\varphi$ 
and viewing angle $\zeta$ \citep{dyksphd}. Pulsar emission spectrum for selected $\zeta$ is obtained through
integrating strip of the photon map $\Delta \zeta$ wide over the pulsar rotation phase.} and spectrum of electrons 
ejected from pulsar magnetosphere (Sect.~\ref{elmsp-char}) were obtained.

\subsection{Spectra of electrons ejected from millisecond pulsar magnetosphere}
\label{elmsp-char}

\begin{figure}
	\begin{center}

	\begin{tabular}{ccc}
	\hspace{-1.2cm}\includegraphics[scale=0.27]{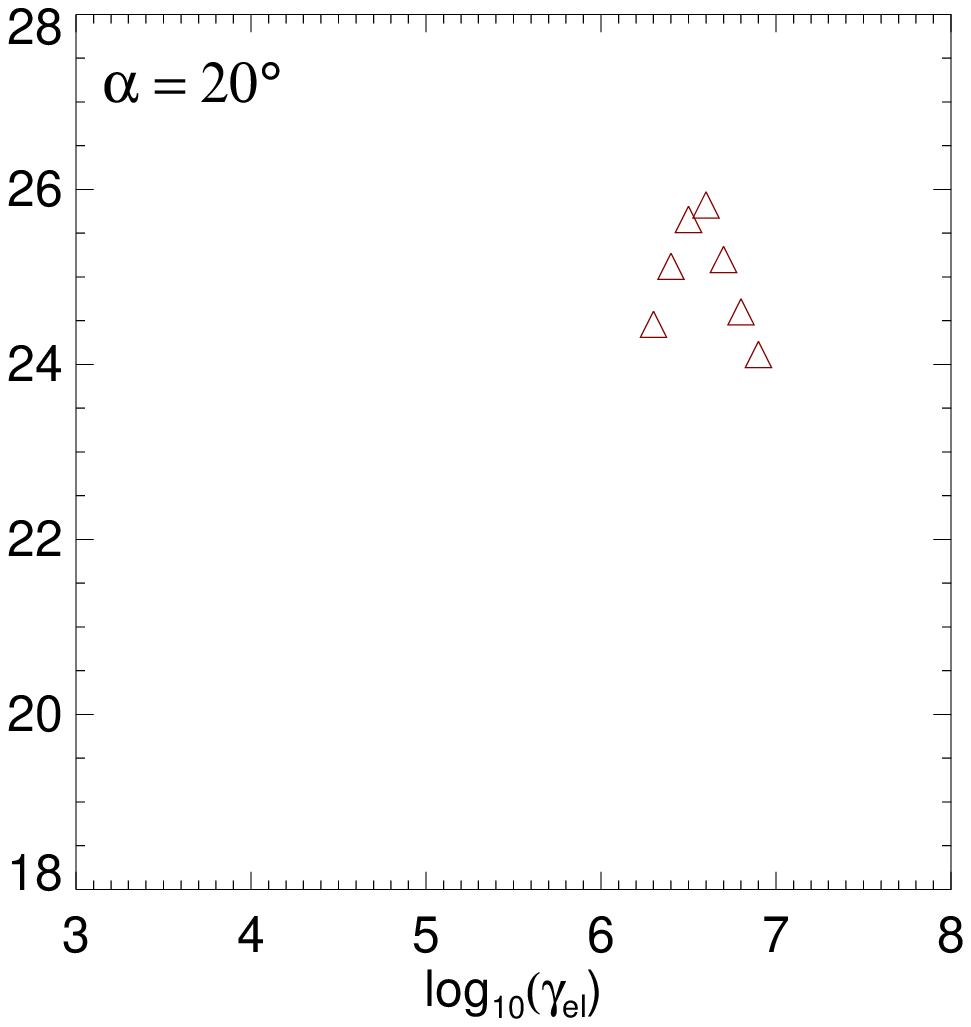}&
	\hspace{-2.3cm}\includegraphics[scale=0.27]{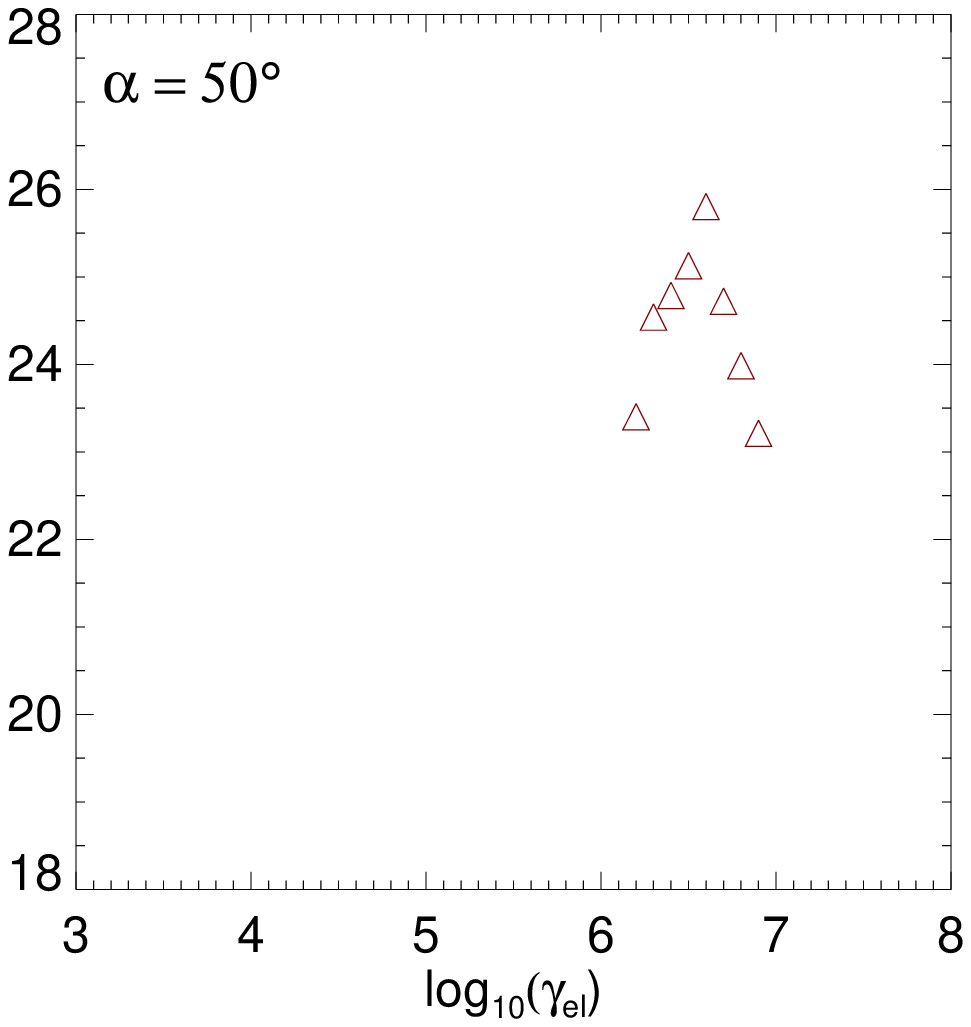}&
	\hspace{-2.3cm}\includegraphics[scale=0.27]{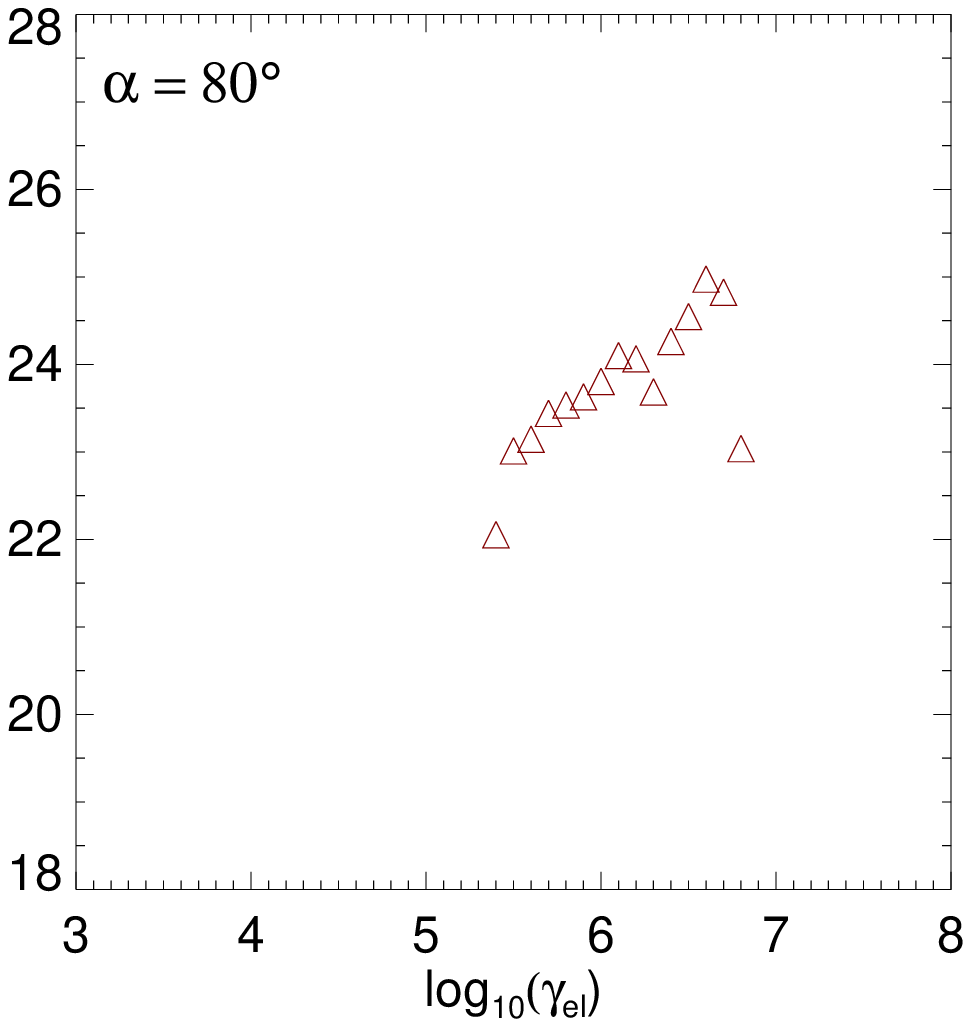}\\
	\hspace{-1.2cm}\includegraphics[scale=0.27]{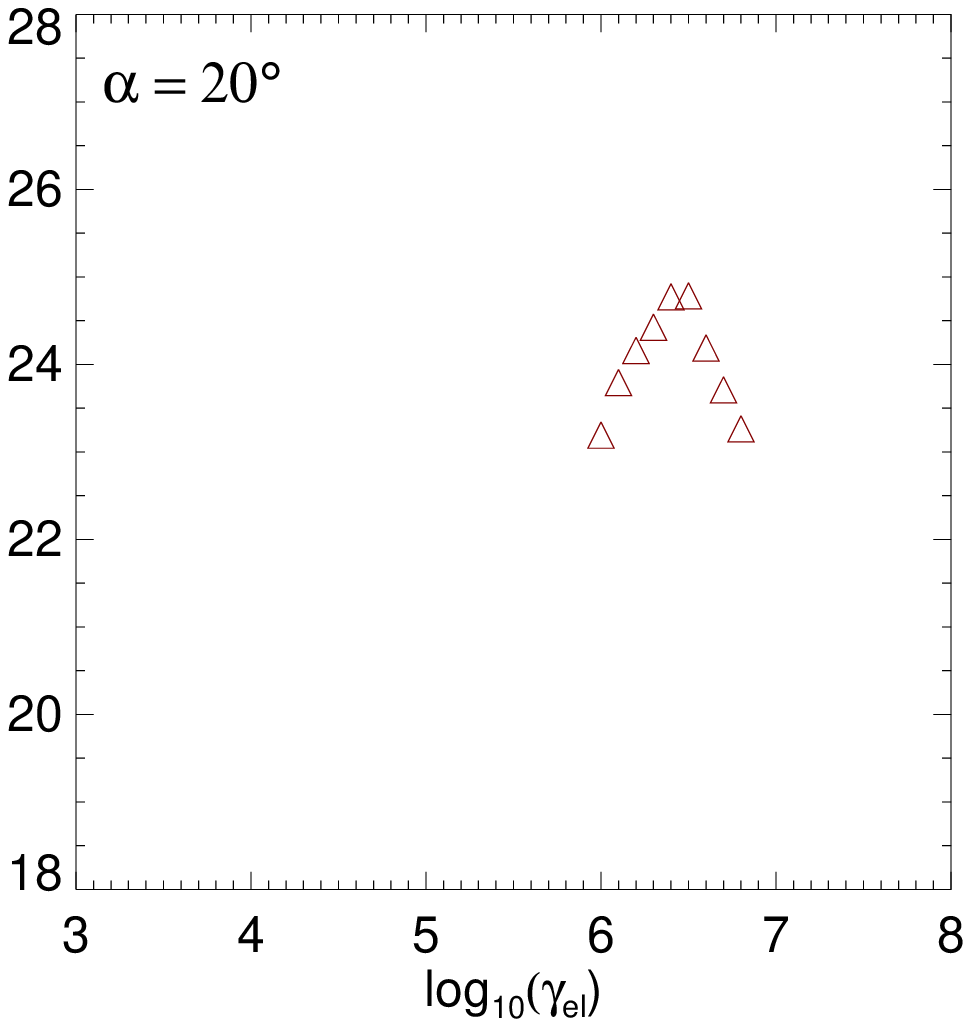}&
	\hspace{-2.3cm}\includegraphics[scale=0.27]{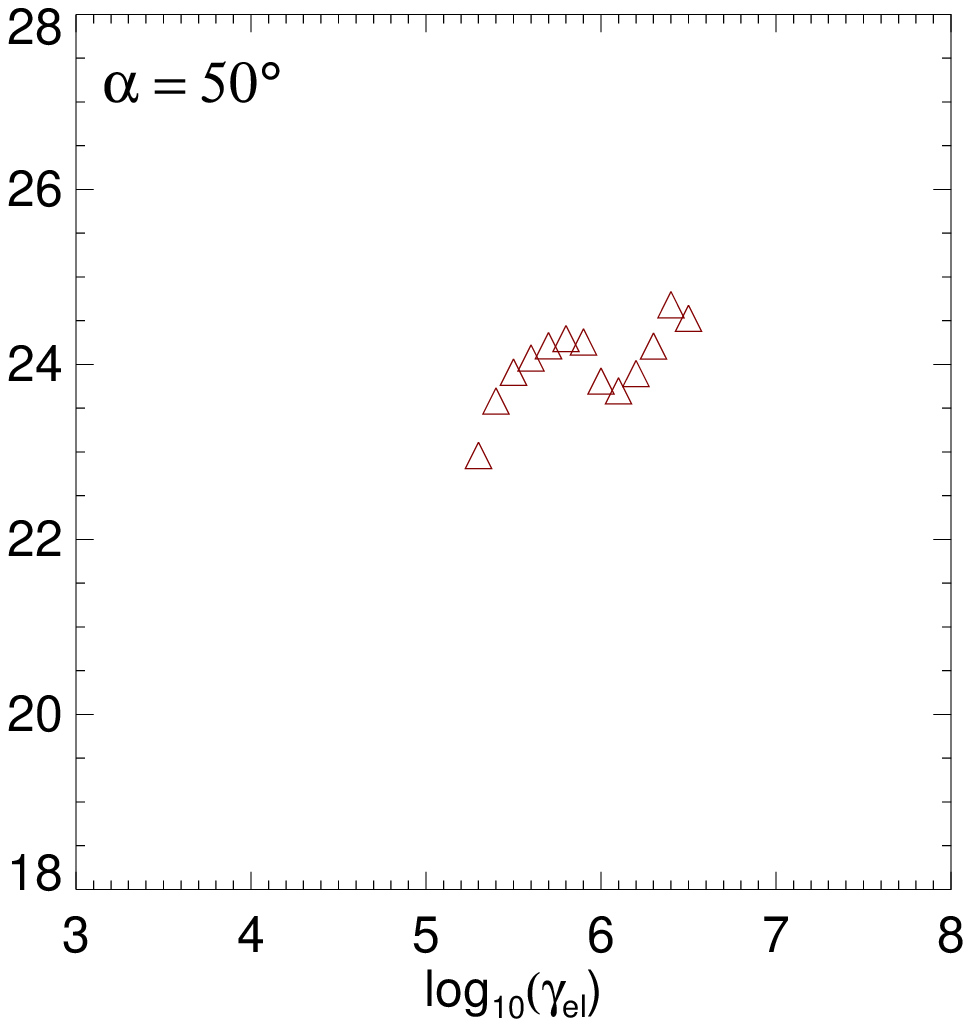}&
	\hspace{-2.3cm}\includegraphics[scale=0.27]{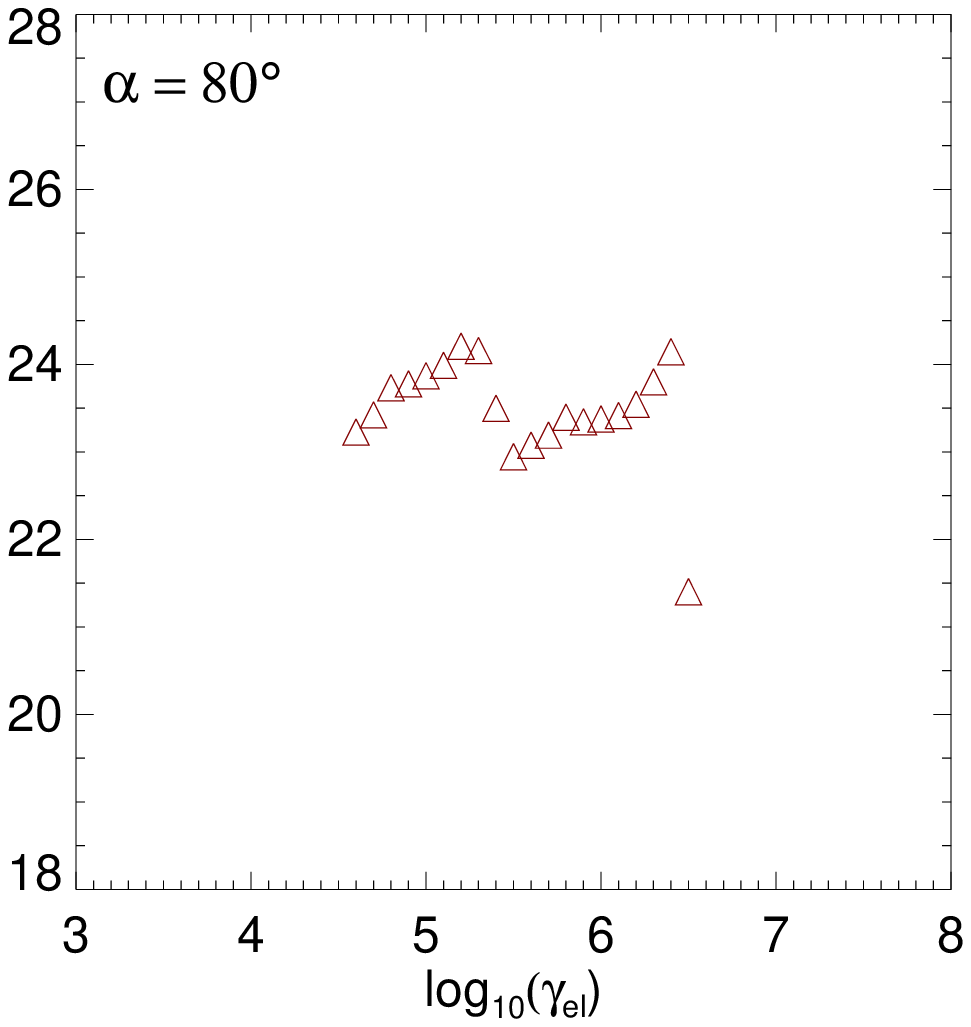}\\
	\hspace{-1.2cm}\includegraphics[scale=0.27]{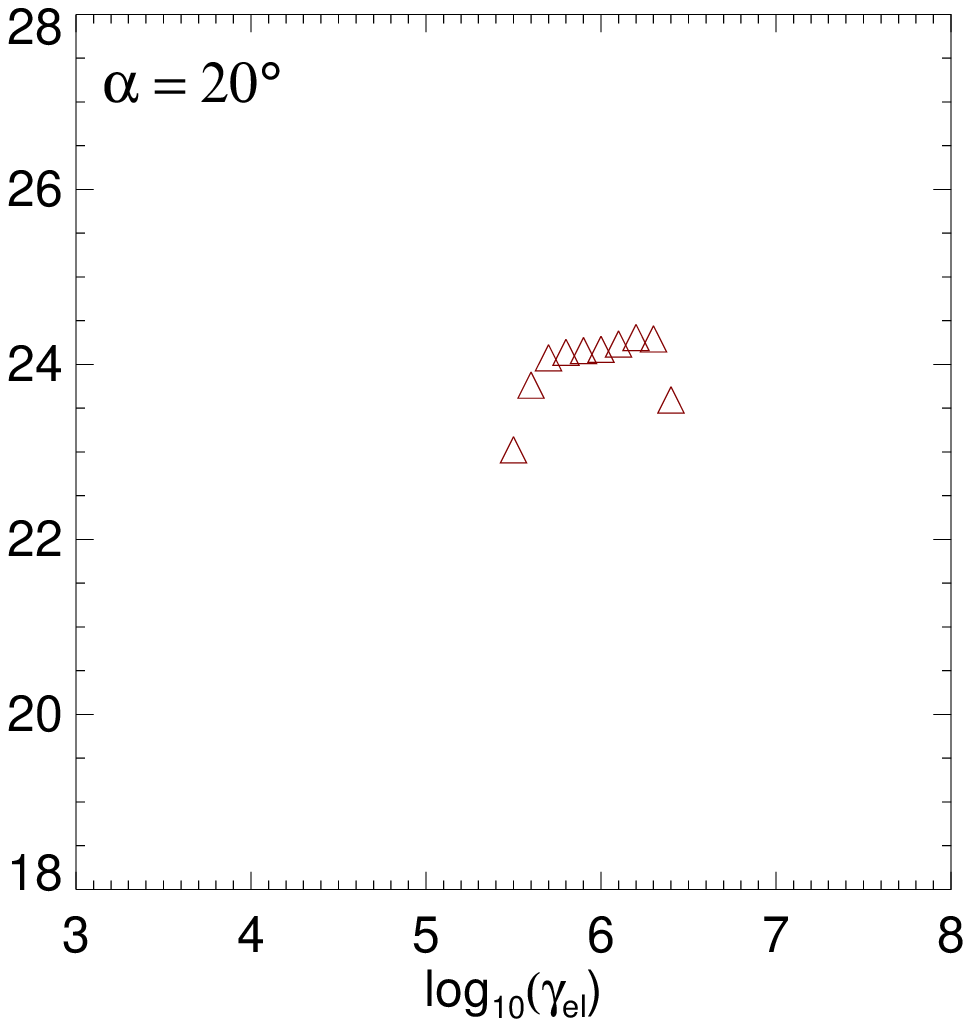}&
	\hspace{-2.3cm}\includegraphics[scale=0.27]{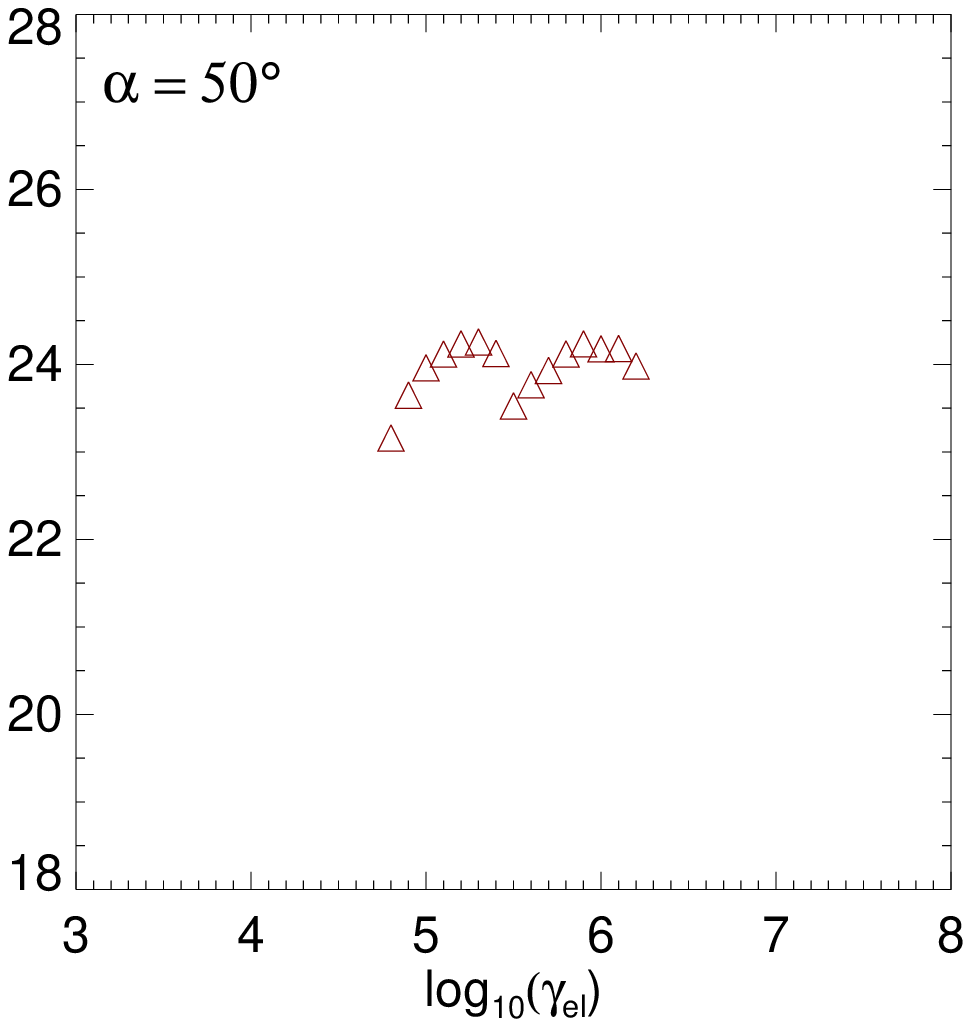}&
	\hspace{-2.3cm}\includegraphics[scale=0.27]{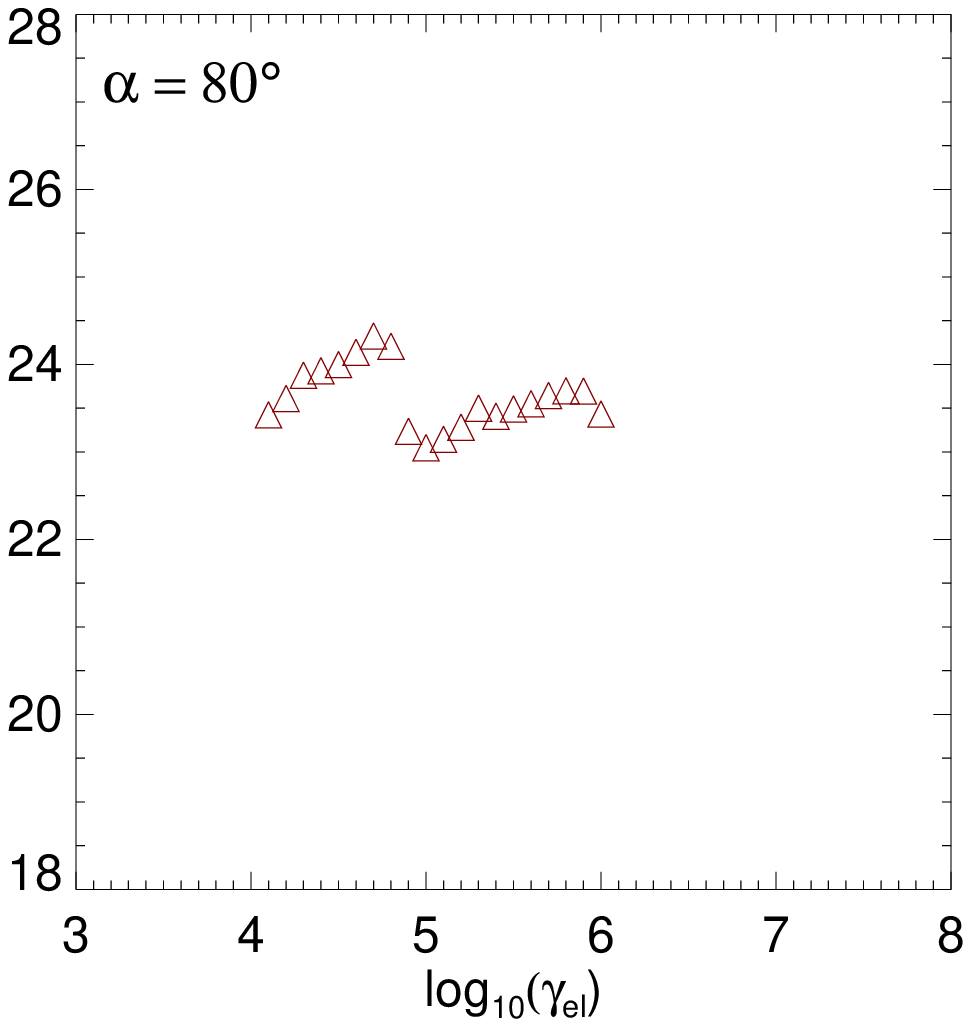}\\
	\end{tabular}
	
	\caption{\small Spectra of electrons escaping through light cylinder for the model with $B_{\mathrm{s}} = 10^{8}$~G.
	The \emph{Y}-axis in each of the subfigures shows $\log_{10}(d\dot{\mathcal{N}}_{\mathrm{el}}/dE_{\mathrm{el}} 
	~\mathrm{[s}^{-1} \mathrm{MeV}^{-1} \mathrm{]})$.
	Spin period $P$ of 1.5, 5.0 and 9.5~ms is presented in the \emph{top}, \emph{middle} and the \emph{bottom} row, 
	respectively. Inclination of the pulsar $\alpha$ changes from $20^{\circ}$, $50^{\circ}$ to $80^{\circ}$ from 
	the \emph{left} column to the \emph{right} one.}
	\label{electron-char-code1}
	\end{center}
\end{figure}

In the calculations we also collect information on the Lorentz factors $\gamma_{\mathrm{el}}$ of the electrons 
escaping the millisecond pulsar magnetosphere. 
Examples of the electron ejection spectra in the form $\log_{10}(d\dot{\mathcal{N}}_{\mathrm{el}}/dE_{\mathrm{el}})$ 
versus $\log_{10} \gamma_{\mathrm{el}}$ are presented in Figure \ref{electron-char-code1}.
The $d\dot{\mathcal{N}}_{\mathrm{el}}$ is the number of primary electrons of 
given energy $E_{\mathrm{el}} = \gamma_{\rm el} ~m_{\rm el}c^{2}$ ($m_{\rm el}$ is the mass of the electron) 
escaping pulsar magnetosphere per unit time. For the case of $B_{\mathrm{s}} = 10^{8}$~G and $\alpha = 20^{\circ}$ 
the electron ejection spectra are single-peaked (the \emph{left} column of Fig.~\ref{electron-char-code1}). 
The Lorentz factors of the particles contributing to the peak are in the range $\sim 10^{5.5} - 10^{7}$. 
For these cases, with the increase of the spin period from 1.5~ms to 9.5~ms the spectral peak becomes 
flatter and shifts towards the lower values of $\gamma_{\mathrm{el}}$. For $P=1.5$~ms the peak centre is at 
$\gamma_{\rm el} \sim 10^{6.6}$, while for the case of $P=9.5$~ms it is at $\gamma_{\rm el} \sim 10^{6.0}$. 
With the increase of the inclination a pronounced low energy tail starts to develop, which eventually 
for slowly rotating synthetic millisecond pulsars (e.g. $P = 5$ ms and 9.5 ms) evolves to form a separate 
low energy peak. Such double-peaked electron ejection spectrum can be seen for the case of $P = 9.5$~ms
and $\alpha = 50^{\circ}$ (the \emph{middle} plot in the \emph{bottom} row of Fig.~\ref{electron-char-code1}). 
Similarly as for the cases with $\alpha = 20^{\circ}$, with the increasing $P$ the shift of the whole 
spectrum towards the lower energies is observed for the cases with $\alpha = 50^{\circ}$ and $80^{\circ}$.
For the model with $P = 9.5$~ms and $\alpha = 80^{\circ}$ this results in the fact that the spectrum spans 
the range of $\gamma_{\mathrm{el}}$ between $\sim 10^{4}$ and $\sim 10^{6}$.
The similar behaviour in the electron spectra is also observed for the models with larger values of 
the magnetic field \citep{zajczykphd}. However, the development of the low energy tail followed by 
the emergence of the low energy peak takes place much later in terms of the pulsar inclination 
(higher $\alpha$) and spin period (larger $P$) than for the low $B_{\mathrm{s}}$ pulsars.

The overall behaviour of the electron ejection spectra is linked to the electric field structure $E_{||}$
in the pulsar magnetosphere (see Sect.~\ref{eacc}). For a given case with $P$ and $B_{\mathrm{s}}$ 
the development of the low energy peak in the electron spectrum starts when more $E_{||}$ solutions 
characterised by the low values of the $\eta_{c}$ parameter are found 
across the magnetosphere. This in turn results in the fact that primary electrons accelerated in different 
parts of the pulsar magnetosphere reach progressively lower energies at the light cylinder. 
When the electric field structure (determined by the value of $\eta_{c}$) evolves into the state when within 
approximately half of the magnetosphere high $\eta_{c}$ electric field solutions are found, and the other
half is characterised by the very low values of $\eta_{c}$ or no solutions at all, then the double-peaked
characteristics of the electron ejection spectrum sets in. Such situation happens exactly when the inclination
becomes $> 40^{\circ}$ (Fig.~\ref{efield-etac-dist1}).
For a given case with $B_{\mathrm{s}}$ and $\alpha$ as the pulsar period increases the magnitude of 
the electric field decreases. Thus, on average the primary electrons reach smaller energies at the light 
cylinder and the whole ejection spectrum moves towards lower values of $\gamma_{\mathrm{el}}$.
Finally, when the pulsar magnetic field strength is increased, the average magnitude of the electric field
in the pulsar magnetosphere is increased. The previously described double-peaked spectral behaviour is
also present for high $B_{\mathrm{s}}$ cases. However, it becomes pronounced only for slowly rotating ($P > 5$ ms)
and highly inclined ($\alpha > 50^{\circ}$) synthetic pulsars because of the much larger average magnitude 
of the electric field. For the same reason, the spectrum shift towards lower electron energies is far less 
severe for the high magnetic field cases ($B_{\rm s} = 9 \times 10^{8}$~G), where the low energy end 
of the spectrum shifts only to $\gamma_{\mathrm{el}} \sim 10^{5}$. For the low magnetic field cases
($B_{\rm s} = 10^{8}$~G) the low energy end of the ejection spectrum can shift even to $\gamma_{\mathrm{el}} 
\sim 10^{4}$.

\section{Gamma-ray emission of synthetic globular cluster}
\label{synth-gc}

Globular clusters have recently been established as a source of the $\gamma$-ray radiation (see Sect.~\ref{gcs-intro}).
Their high energy emission is attributed to the ensemble of millisecond pulsars residing in their cores. Combined
magnetospheric activity of the millisecond pulsar population is thought to be the source
of the cluster emission above 100~MeV. The spectral shapes of the radiation observed for GCs \citep{abdo09, abdo10, 
kong10, tam11} seem to confirm this scenario. The relativistic electrons injected into the cluster environment
by the MSPs can up scatter the ambient photon fields (cosmic microwave background, infrared background, and
starlight either originating from stellar population in the globular cluster or from stars in the Galaxy). This
inverse Compton scattering scenario has been proposed to explain the GCs $\gamma$-ray emission
in the GeV and TeV energy range \citep{bednarek07, cheng10, venter09}. The observational properties of the $\gamma$-ray
emission of GCs cannot be explained solely with the magnetospheric activity scenario or the ICS scenario. Thus,
most probably the interplay between the two models gives rise to the overall GeV and TeV emission of the globular
clusters (see Sect.~\ref{gcs-intro} for more details).

\subsection{Numerical scheme of the synthetic globular cluster}

In order to simulate the $\gamma$-ray emission of a synthetic globular cluster both the magnetospheric activity 
of the population of millisecond pulsars residing in the cluster, and also the ICS scenario is taken into account. 
The magnetospheric component (Sect.~\ref{magnet-comp}) is constructed using the results of the \emph{pair 
starved polar cap} model calculations for different cases of millisecond pulsars (Sect.~\ref{eacc}).
The ICS component is calculated in two steps, first a cumulative spectrum of electrons injected by ensemble 
of synthetic millisecond pulsars into the cluster environment is computed. The cumulative electron spectrum 
is constructed using the electron ejection spectra calculated with the \emph{pair starved polar cap
model} (see Sects.~\ref{elmsp-char}). Later, the cumulative spectrum is fed into the numerical model of the cluster 
(BS07) to produce the ICS radiation (Sect.~\ref{ics-comp}).

The numerical procedure that calculates the $\gamma$-ray spectrum originating from the ensemble of millisecond 
pulsars residing in the core of the synthetic cluster, and also the cumulative spectrum of electrons ejected 
from the magnetospheres of those synthetic MSPs is written in the IDL language.

It is assumed that the number of millisecond pulsars residing in the synthetic cluster is $N_{\rm PSR}$. These
pulsars reside in the cluster core, which in the numerical calculations is resolved assuming that these pulsars
are concentrated in one point placed in the cluster centre. In the first step, the basic parameters 
(a spin period $P$, a magnetic field strength $B_{\rm s}$ and an inclination angle $\alpha$) are selected for each 
of the millisecond pulsars residing in the modelled cluster. 
The values of these parameters are randomly selected from the discrete distribution determined by the synthetic 
MSP population in the database (Sect.~\ref{eacc}). The random selection is performed with the IDL function 
\texttt{randomu} \citep{idlrandom}.

\subsection{Magnetospheric emission of the population of millisecond pulsars in the cluster}
\label{magnet-comp}

Having selected basic parameters ($P$, $B_{\rm s}$, $\alpha$) that characterise each of $N_{\rm PSR}$
millisecond pulsars populating our synthetic globular cluster, we are ready to compute the magnetospheric 
contribution of the ensemble of MSPs to the high energy emission of the synthetic cluster. 
Each of the selected MSPs emits high energy photons originating in the radiation processes that take 
place in the pulsar magnetospheres. As already discussed in Sect.~\ref{msp-model}, in the PSPC model 
only curvature photons are emitted.
\textbf{ Within the framework of the studied model and the selected range of pulsar parameters 
($P$, $B_{\rm s}$, $\alpha$) there is no additional contribution (with respect to the curvature radiation) 
at the GeV energy range from synchrotron radiation due to secondary pairs \citep[see e.g.,][]{rudak99}.
Moreover, any synchrotron radiation due to cyclotron resonant absorption of radio waves is beyond 
the scope of the studied model.}

To create the cumulative magnetospheric contribution from the millisecond pulsars, first a viewing angle
$\zeta$ had to be selected for each pulsar. Its value is randomly selected from the $\sin \zeta$ 
distribution. Having selected the viewing angle, the observed $\gamma$-ray 
spectrum is constructed for each millisecond pulsar populating the modelled cluster:
\begin{equation}
	F_{\rm obs}(\zeta; \mathcal{E}) = \frac{1}{2\pi D^{2}}
	\int_{0}^{2 \pi} \mathcal{E}\:
	\frac{d\dot{N}_{\gamma}}{d\mathcal{E}d\Omega}\: d\varphi  ~,
\end{equation}
where $\zeta$ is the randomly selected viewing angle, $\varphi$ is the pulsar rotation phase and $D$ 
is a distance to the simulated cluster. This in turn allows for computing the overall millisecond 
pulsar magnetospheric contribution to the observed $\gamma$-ray emission of the synthetic cluster:
\begin{equation}
	F_{\rm N_{PSR}}(\mathcal{E}) = \sum_{k=1}^{N_{\rm PSR}} F_{\mathrm{obs}, k}(\zeta; \mathcal{E}) ~.
	\label{gc-heflux}
\end{equation}

\begin{figure}
	\begin{center}
	\begin{tabular}{c}
	\hspace{-0.25cm}\includegraphics[scale=0.48]{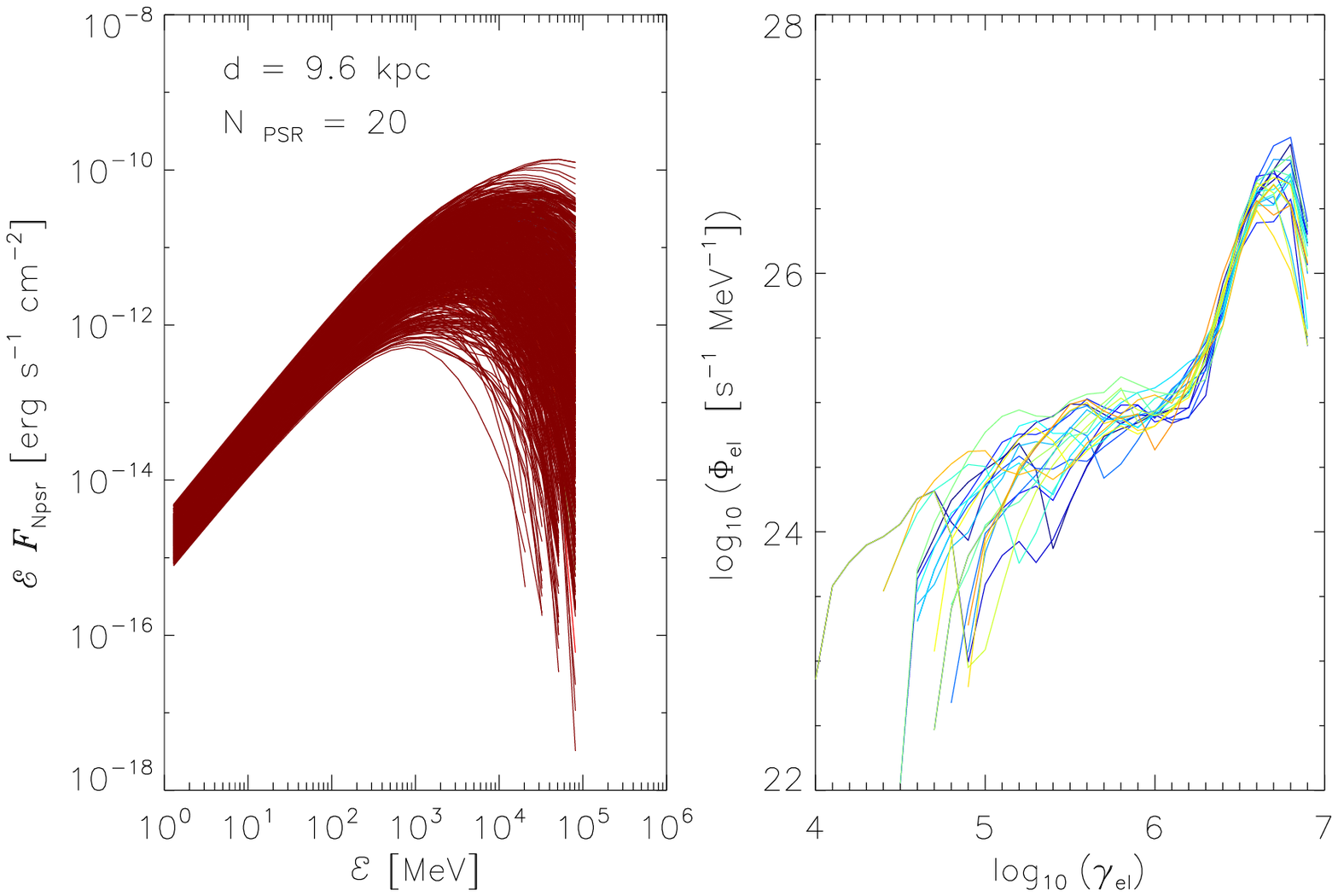}\\
	\hspace{-0.25cm}\includegraphics[scale=0.48]{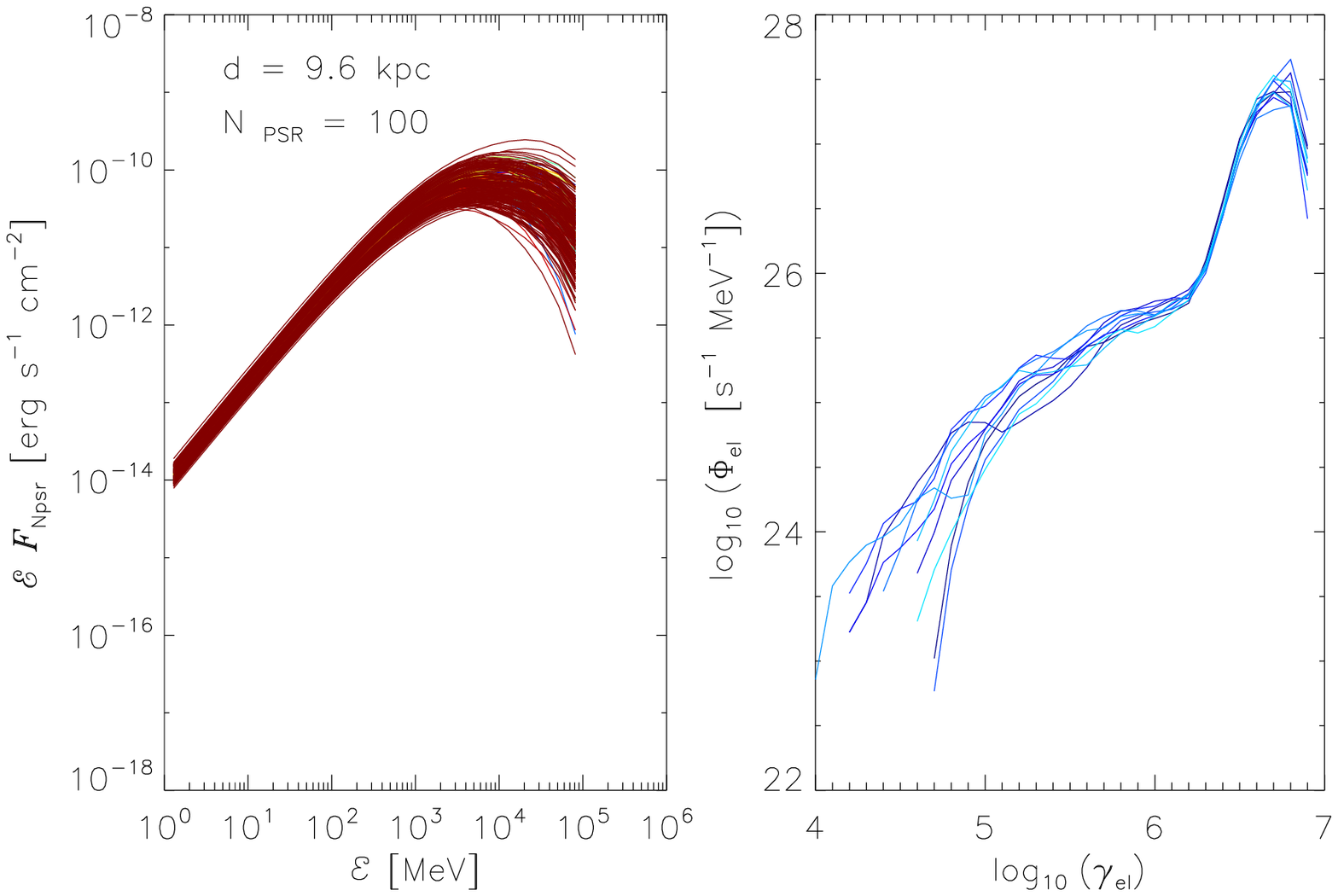}\\
	\end{tabular}

	\caption{\small The magnetospheric $\gamma$-ray spectra (\emph{left} column) and the cumulative 
	electron spectra (\emph{right} column) computed for different realisations of synthetic globular 
	cluster. The magnetospheric $\gamma$-ray spectra are presented in $\mathcal{E} F_{\rm N_{\rm PSR}}(\mathcal{E})$
	form (i.e. a $\gamma$-ray flux described by Eq.~\ref{gc-heflux} multiplied by the photon energy $\mathcal{E}$).
	The cumulative electron spectra are given by Eq.~\ref{elspec-tot}. In the \emph{top} 
	row a modelled GC harbouring 20 millisecond	pulsars ($N_{\rm PSR}$) is presented, while in 
	the \emph{bottom} row a model with $N_{\rm PSR} = 100$ is shown. In both cases, the distance to 
	the cluster is taken to be $d = 9.6$~kpc.}
	\label{synth-gc-example}
	\end{center}
\end{figure}

In the \emph{left} column of Fig.~\ref{synth-gc-example} examples of the cluster $\gamma$-ray spectra resulting 
\emph{only} from the magnetospheric emission of population of millisecond pulsars are presented.
In the \emph{top left} panel of Fig.~\ref{synth-gc-example} one thousand cluster $\gamma$-ray spectra
in the form of $\mathcal{E} F_{\rm N_{\rm PSR}}(\mathcal{E})$ are shown. It is assumed that 20 millisecond 
pulsars reside in the modelled cluster. The distance to the cluster is 9.6~kpc. The obtained spectral characteristics 
of the synthetic cluster is diverse both in terms of the photon cut-off energy and the maximum flux density level.
The energy of the cut-off $\mathcal{E}_{\rm cut}$ falls in the range between $10^{3}$~MeV and $\sim 5 
\times 10^{4}$~MeV. The cluster spectra characterised by the lowest value of $\mathcal{E}_{\rm cut}$
are also the spectra with the lowest maximum flux density level $\sim 10^{-12}$~erg$^{-1}$~s$^{-1}$~cm$^{-2}$.
The highest generated maximum flux density level is $\sim 10^{-10}$~erg$^{-1}$~s$^{-1}$~cm$^{-2}$.
Assuming that there are only 20 pulsars in the cluster we are able to ge\-ne\-ra\-te the magnetospheric
$\gamma$-ray spectra of the globular cluster that differ by two orders of magnitude in terms of
the flux density level.
In the \emph{bottom left} panel of Fig.~\ref{synth-gc-example} three hundred cluster $\gamma$-ray spectra
are displayed. The simulated globular cluster harbours $N_{\rm PSR} = 100$ millisecond pulsars, and
is at the distance of 9.6 kpc. The cut-off energies of the simulated magnetospheric cluster spectra
span the energy range between $\sim 2 \times 10^{3}$~MeV and $\sim 2 \times 10^{4}$~MeV. The maximum
level of the flux density varies between $\sim 3 \times 10^{-11}$~erg$^{-1}$~s$^{-1}$~cm$^{-2}$
and $\sim 3 \times 10^{-10}$~erg$^{-1}$~s$^{-1}$~cm$^{-2}$.
The comparison of the models presented in the \emph{top left} ($N_{\rm PSR} = 20$) and in 
the \emph{bottom left} ($N_{\rm PSR} = 100$) panel of Fig.~\ref{synth-gc-example} shows that
the synthetic globular clusters harbouring smaller number of millisecond pulsars can produce similar 
magnetospheric $\gamma$-ray spectra (in terms of the cut-off energy and the flux density level)
as the ones produce by the clusters with larger pulsar population.

\subsection{ICS component to the $\gamma$-ray emission of the synthetic globular cluster}
\label{ics-comp}

The first step in obtaining an ICS emission component to the $\gamma$-ray radiation
from globular clusters is creation of a cumulative spectrum of electrons injected by the population
of synthetic MSPs into the GC environment. The electron ejection spectrum (see Sect.~\ref{elmsp-char}) for a given 
pulsar ($P$, $B_{\rm s}$, $\alpha$) is independent of a viewing angle, so once we have selected
these basic pulsar parameters for all $N_{\rm PSR}$ pulsars in the synthetic globular cluster, 
we construct the electron ejection spectrum for each synthetic pulsar:
\begin{equation}
	\Phi_{\mathrm{el}, k}(\gamma_{\rm el}) = \frac{d\dot{\mathcal{N}}_{\rm el}(k)}{d E_{\rm el}} ~,
\end{equation}
where $\dot{\mathcal{N}}_{\rm el}$ is the number of electrons with the energy $E_{\rm el} = \gamma_{\rm el} 
~m_{\rm el}c^{2}$ ejected from pulsar magnetosphere per unit time. Then to obtain the cumulative spectrum 
of electrons, the electron ejection spectra from all the pulsars in the modelled globular cluster are
summed up:
\begin{equation}
	\Phi_{\rm el} = \sum_{k=1}^{N_{\rm PSR}} \Phi_{\mathrm{el}, k}(\gamma_{\rm el}) ~.
	\label{elspec-tot}
\end{equation}

Examples of the cumulative electron spectra obtained for the population of millisecond pulsars 
residing in the synthetic globular cluster are presented in the \emph{right} column of Fig.~\ref{synth-gc-example}.
The results for the cluster harbouring 20 synthetic millisecond pulsars is presented in the
\emph{top right} panel of Fig.~\ref{synth-gc-example}. In the plot there are 25 different cumulative 
electron spectra $\Phi_{\rm el}$ displayed. A common feature present in all of the spectra is a narrow 
peak positioned at high electron Lorentz factors. Its centre is around $\gamma_{\rm el} \simeq 10^{6.5}$. 
A low energy tail is also characteristic of all the presented $\Phi_{\rm el}$ spectra. It stretches down 
to electron Lorentz factors $\sim 10^{4}$. On average the particle flux level in the tail is two orders 
of magnitude lower than in the peak ($\Phi_{\rm el} \sim 10^{25}$~s$^{-1}$~MeV$^{-1}$ in the tail versus 
$\Phi_{\rm el} \sim 10^{27}$~s$^{-1}$~MeV$^{-1}$ in the peak). The shape of the tail is different between 
different globular cluster simulations. There are spectra where the tail is characterised by a type 
of plateau going from $\gamma_{\rm el} \sim 10^{5}$ to $10^{6}$, which is followed by a sharp decline 
towards lower Lorentz factor values. There are also cases where the low energy tail declines steadily 
towards small $\gamma_{\rm el}$ values from the point it emerges from the high energy peak. 
The cumulative electron spectra $\Phi_{\rm el}$ obtained for the synthetic globular cluster harbouring
100 millisecond pulsars are presented in the \emph{bottom right} panel of Fig.~\ref{synth-gc-example}.
There are 10 different spectra $\Phi_{\rm el}$ presented in the plot. Similarly as for the case of
the cluster harbouring 20 pulsars, the narrow high energy peak is present in all of the simulated
electron spectra $\Phi_{\rm el}$. It is centred at $\gamma_{\rm el} \simeq 10^{6.5}$, and the maximum
value of the particle flux is $\sim 10^{27.5}$~s$^{-1}$~MeV$^{-1}$. The average flux level in the low
energy tail is by two orders of magnitude lower than in the peak. The shapes of the low energy tail
are less diverse with respect to the case of the synthetic cluster with smaller pulsar population.
In all simulated cases the tail declines steadily towards low $\gamma_{\rm el}$ values right from 
the point it emerges from the high energy peak. The particle flux level in the tail changes from
$\sim 10^{26}$~s$^{-1}$~MeV$^{-1}$ at its high energy end ($\gamma_{\rm el} \sim 10^{6.5}$) down to
$\sim 10^{23}$~s$^{-1}$~MeV$^{-1}$ at its low energy end ($\gamma_{\rm el}$ between $\sim 10^{4}$ 
and $\sim 10^{5}$).

In order to simulate the inverse Compton scattering component to the $\gamma$-ray spectrum of the synthetic 
globular cluster, we are using a numerical model of the cluster developed by BS07. 
Because millisecond pulsars are concentrated in the cluster core, it is assumed in 
the model that the relativistic electrons (their distribution is given by Eq.~\ref{elspec-tot}) are injected 
in the centre of the globular cluster, and further they diffuse gradually towards outer parts of the cluster. 
The diffusion of particles is treated in the Bohm diffusion approximation. On their outward way, electrons 
interact with the cluster magnetic field, and also with ambient photon fields. 
BS07 (see fig.~1) showed that the radiation field within the GC and its nearby surrounding seems 
to be quite homogeneous. At the same time the electron diffusion distance mildly increases when moving 
towards the GC outskirts. Simple calculations show that the average diffusion distance for electrons injected 
in the cluster core and at its edge differ only by a factor of 2. Therefore the effect of distribution of pulsars 
within the core of globular cluster on the cluster's ICS emission is expected not to be very significant\footnote{
From the selected number of millisecond pulsars used to model synthetic $\gamma$-ray spectra of 47 Tuc 
and Ter 5 (see Sect.~\ref{comp-sim-obs}) only 5 out of 23 and 1 out of 31 (http://www.naic.edu/~pfreire/GCpsr.html), respectively, 
lie beyond the distance of $2\times R_{\rm c}$ but within the radius of $3\times R_{\rm c}$ ($R_{\rm c}$ is the cluster core radius). 
This is relatively small number. Thus, assuming that all MSPs are situated in the cluster core does not cause the ICS emission to be 
overestimated.} (taking into account the uncertainties of the discussed scenario).

Previously in their calculations, BS07 did not take into account synchrotron energy losses 
suffered by particles propagating through the cluster. In the presented calculations synchrotron losses
are computed accordingly. However, the produced synchrotron radiation does not contribute to overall 
$\gamma$-ray emission of the cluster.
The characteristic energies of the synchrotron photons produced by electrons escaping from 
the pulsar magnetospheres and propagating within the globular cluster are far below the $\gamma$-ray 
range. They can be estimated from $\varepsilon_{\rm char} \approx m_{\rm el} c^{2} (B_{\rm GC}/B_{\rm crit})\gamma_{\rm el}^2$,
where $B_{\rm crit}$ is the critical magnetic field \citep[see e.g.,][]{thompson95}.
For the electrons with Lorentz factors of the order of $3\times 10^6$ and the magnetic field strengths 
within GC of the order of 10 $\mu$G, the synchrotron photons have characteristic energies in the optical 
range, i.e. they will be very difficult to observe due to the huge background from the GC stars.

In the calculations a following set of the globular cluster magnetic fields $B_{\rm GC} = (0.3$, $1.0$, 
$3.0$, $10$, $30)$~$\mu$G is taken into account.
A main source of photons permeating the cluster are stars residing in the cluster itself. The stellar 
photons can be up-scattered by the relativistic electrons (with energies up to $\sim 1$~TeV) via inverse Compton 
scattering up to TeV energies. The density of the stellar radiation field is not constant throughout the cluster, 
but it should mimic the distribution of stars in the system. For this reason, the density of stars in the cluster 
is described in the framework of the Michie-King model \citep{michie63}; this allows for determination of the energy 
density of stellar photons $U_{\rm stellar}$ as a function of the distance from the cluster centre. In order 
to determine the energy density of the stellar photons within the cluster information on the cluster visual 
luminosity $L_{\rm VIS}$ is necessary. Moreover, the core radius $R_{\rm c}$, the half-mass radius $R_{\rm h}$ 
and the tidal radius $R_{\rm t}$ are also necessary for determining $U_{\rm stellar}$ in the cluster
\citep[see eqs.~4 and 5 of][]{bednarek07}. 

As shown by BS07, the energy density of the stellar radiation field ($U_{\rm stellar} \approx 
300$~eV cm$^{-3}$ inside a cluster core for the case where core radius is 0.5~pc and cluster luminosity is 
at the level of $10^{5} L_{\odot}$) dominates over the energy density of the cosmic microwave 
background (CMB; $U_{\rm CMB} = 0.25$~eV cm$^{-3}$). However, the latter photon field is important for estimating 
the ICS component (from tens MeV up to hundreds GeV) produced on the high energy electrons (energies above $\sim 
1$~TeV). The described processes are simulated using the Monte Carlo method, and lead to the production of the ICS 
emission spectrum of the globular cluster. The details of the numerical code are presented in the work of 
BS07.

The results of the numerical simulations performed for two globular clusters: Terzan 5 and 47 Tucanae are
presented in Sect.~\ref{comp-sim-obs}. The outcome of the simulations is the total $\gamma$-ray spectrum
for each of the clusters. The spectrum consists of the magnetospheric and ICS component.

\section{Comparison of the simulations with the observations}
\label{comp-sim-obs}

Using the numerical procedure presented in Sect.~\ref{synth-gc}, $\gamma$-ray spectra were computed for two
selected globular clusters: Terzan 5 and 47 Tucanae. Where possible, the simulated spectra are compared
with the publicly available observational data.

An important comment about the selection of pulsar periods for the population of MSPs residing in the
clusters is necessary. Because for each of the selected clusters a number of millisecond pulsars was 
detected \citep[see e.g.,][]{ransom05, manchester91, camilo00} and their spin periods were determined, 
a different approach to the problem of the spin period $P$ selection than the one presented in 
Sect.~\ref{synth-gc} was used.
From the list of the known MSPs in each cluster, pulsars with periods satisfying the criteria $P \lesssim
11$~ms were selected as the sole members of the population of the millisecond pulsars in the cluster. 
Knowing their spin periods and the density of the simulated grid in our database (Sect.~\ref{eacc}), 
for each MSP the spin period value as close as possible to the true $P$ value was assigned from the database.
For each MSP the value of the magnetic field $B_{\rm s}$ and the inclination angle $\alpha$
was randomly selected from the discrete values simulated in our database (Sect.~\ref{eacc}).
From this point onwards, the procedure leading to the production of the cluster $\gamma$-ray spectrum matches 
the one presented in detail in Sect.~\ref{synth-gc}.

\subsection{The case of Terzan 5}
\label{ter5-sim}

Currently, Terzan 5 is the globular cluster with the largest number of the detected millisecond pulsars.
There are 34 MSPs observed in the cluster (see the catalogue of P.~C. Freire\footnote{ 
http://www.naic.edu/$\sim$pfreire/GCpsr.html}). In addition, it is one of the first globular clusters 
detected in $\gamma$-rays by Fermi/LAT \citep{abdo10, kong10}. Its luminosity estimated in the energy 
range between 100~MeV up to around 20~GeV is $L_{\gamma} \simeq 2.6 \times 10^{35}$~erg s$^{-1}$ \citep{abdo10} 
assuming the cluster is situated at a distance of $d \simeq 5.5$~kpc from the Sun. The observed $\gamma$-ray 
spectrum can be well fitted with a power law with an exponential cut-off $N(\mathcal{E}) \sim \mathcal{E}^{-\Gamma}
e^{-\mathcal{E}/\mathcal{E_{\rm c}}}$ ($N(\mathcal{E}) \equiv d\dot{N}_{\gamma}/d\mathcal{E}$ is a photon flux, 
$\Gamma$ is a photon index, and $\mathcal{E_{\rm c}}$ is a cut-off energy).
The spectral parameters estimated for Ter~5 are $\Gamma \simeq 1.4$ and $\mathcal{E_{\rm c}} \simeq 2.6$~GeV
\citep{abdo10}. Recently, the cluster has been detected by the H.E.S.S. telescope \citep{gchess} in the TeV
domain. The observed photon flux in the energy range above 0.4~TeV
is $\simeq 1.2 \times 10^{-12}$~cm$^{-2}$~s$^{-1}$, while the photon index of the spectrum fitted with
the power law $N(\mathcal{E}) \sim \mathcal{E}^{-\Gamma}$ is $\Gamma \simeq 2.5$. The TeV emission detected
from the direction of Ter 5 is slightly shifted with respect to the cluster core and it extends well beyond
the cluster tidal radius. \citet{gchess} speculate that the observed
TeV emission may be a source coincidence, e.g. with a PWN associated with a radio-quiet pulsar, or that the 
TeV $\gamma$-rays could have hadronic origin. If neither of these scenarios is true, then Ter~5 is the only 
globular cluster detected so far in the TeV domain with the Cherenkov telescopes. 

Important parameters of Ter 5 cluster that are used in calculations of the ICS spectral component are the visual 
luminosity of the cluster $L_{\rm VIS} \simeq 1.5 \times 10^{5} L_{\odot}$ \citep{harris96}. The core radius 
$R_{\rm c}$, the half-mass radius $R_{\rm h}$ and the tidal radius $R_{\rm t}$ were taken from \citet{harris96}.
The ambient photon fields up scattered via inverse Compton process are the cosmic microwave background 
($U_{\rm CMB} = 0.25$ eV cm$^{-3}$) and the photons
from the stellar population in the cluster (the energy density was estimated with eqs.~4 and 5 of BS07).

\begin{figure}
	\begin{center}
	\begin{tabular}{c}
	\includegraphics[scale=0.44]{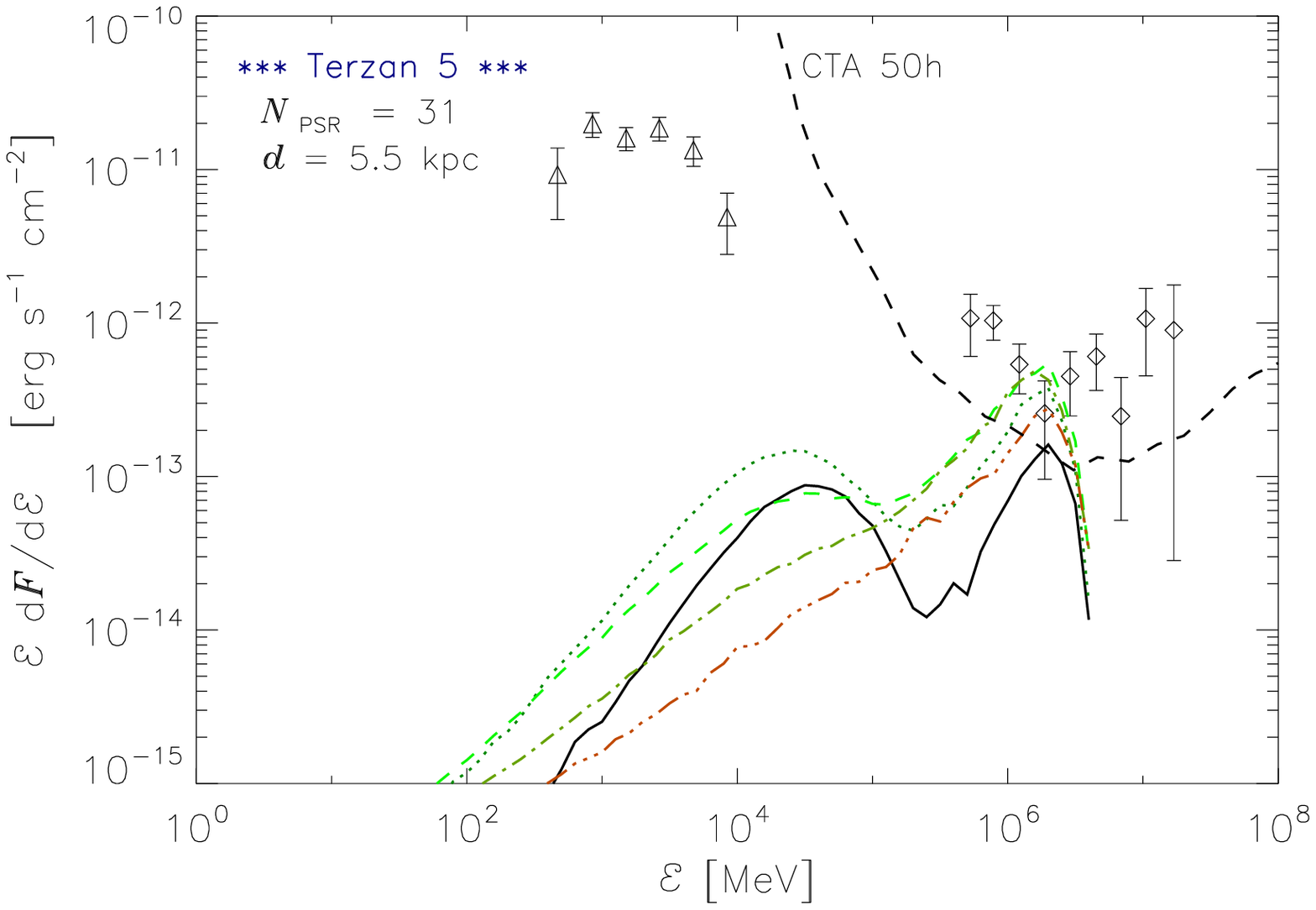}\\
	\includegraphics[scale=0.44]{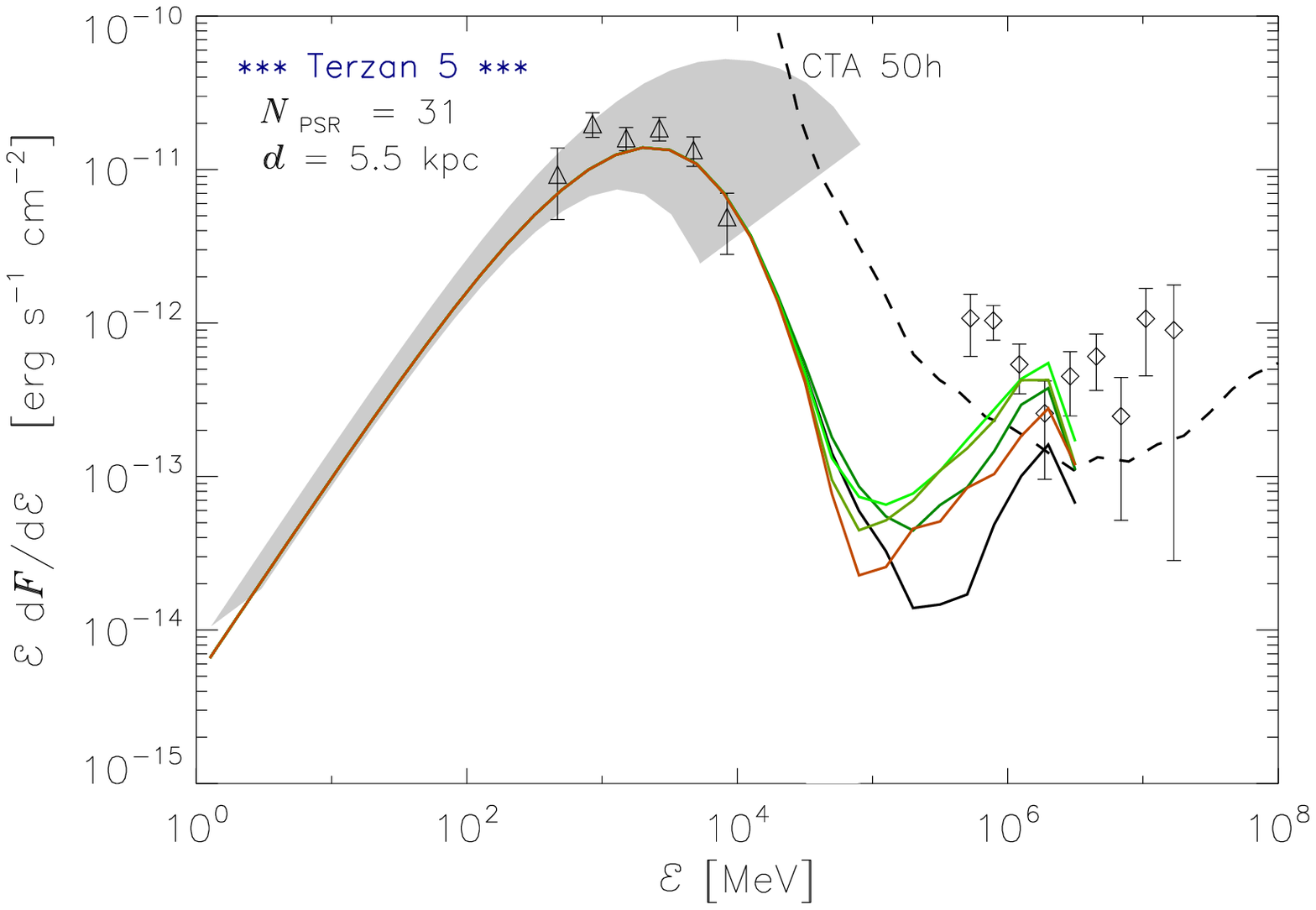}\\
	\end{tabular}

	\caption{\small Synthetic spectra calculated for the globular cluster
	Terzan 5. The total number of millisecond pulsars in the cluster is $N_{\rm PSR} 
	= 31$. The distance to the cluster is assumed to be $d = 5.5$~kpc. \emph{Top 
	panel:} ICS spectra computed for different cluster magnetic field values 
	$B_{\rm GC}$. Solid black, green dotted, green dashed, green dot-dashed and brown
	dot-dot-dashed lines show models with $B_{\rm GC}$ of $0.3$, $1.0$, $3.0$, $10$,
	$30$~$\mu$G, respectively. \emph{Bottom panel:} Total $\gamma$-ray spectra
	(including magnetospheric and ICS component) are depicted with solid lines. 
	Colours of the lines correspond to the value of the magnetic field in the globular
	cluster - from black, through shades of green to brown correspond to values of
	$B_{\rm GC}$ from 0.3~$\mu$G up to 30~$\mu$G. The colour-coding of the lines is 
	the same as for the \emph{top} panel. ICS spectra in the \emph{bottom} panel 
	were re-binned	to match the resolution of the magnetospheric component spectra. 
	Grey area shows the region occupied by the magnetospheric spectra of the cluster
	obtained for presented simulation runs (this can be treated as the uncertainty 
	of the MeV to GeV emission spectrum of GC). Triangles show the $\gamma$-ray 
	spectrum (with uncertainties) of Ter 5 as observed by the Fermi/LAT instrument
	\citep{abdo10}. Diamonds depict spectrum (and its errors) observed from 
	the direction of the cluster by the H.E.S.S. telescope \citep{gchess}. Dashed 
	line shows the CTA differential sensitivity curve with 50~h of integration 
	\citep{cta11}.}
	\label{synth-gc-ter5}
	\end{center}
\end{figure}

Out of 34 known MSPs in the cluster, 3 have spin periods larger than 11~ms. Thus, $N_{\rm PSR} = 31$ were 
used to simulate $\gamma$-ray radiation from the cluster. The results are presented in Fig.~\ref{synth-gc-ter5}. 
For one set of $N_{\rm PSR}$ with chosen ($P$, $B_{\rm s}$, $\alpha$) one cumulative spectrum of electrons
(given by Eq.~\ref{elspec-tot}) is obtained. This spectrum is then injected into the cluster environment 
and it interacts with the ambient photon fields leading to the production of the ICS component of the 
$\gamma$-ray spectrum of the modelled globular cluster. Simultaneously, for one set of $N_{\rm PSR}$ many 
different cumulative magnetospheric spectra (given by Eq.\ref{gc-heflux}) are simulated. The cumulative 
magnetospheric spectrum is the component of the synthetic cluster $\gamma$-ray spectrum stretching from
around MeV up to around few tens of GeV. The grey area in the \emph{bottom} panel of Fig.~\ref{synth-gc-ter5} 
shows the region occupied by the group of different magnetospheric spectral components simulated for 
the cluster. It can be treated as the uncertainty of the GC emission spectrum in the MeV to GeV energy domain. 
In the \emph{bottom} panel of Fig.~\ref{synth-gc-ter5} the simulated total $\gamma$-ray spectra for Ter 5
are presented with solid lines. To construct the total spectra, one magnetospheric component out of 
the simulated group was chosen. Five different ICS components were obtained for the cluster. They were
simulated taking different values of the cluster magnetic field $B_{\rm GC}$ (see Sect.~\ref{ics-comp}).
For clarity, these ICS components are presented in the \emph{top} panel of Fig.~\ref{synth-gc-ter5}
separated from the chosen magnetospheric component.

The overall shape of the ICS spectral component (see the \emph{top} panel of Fig.~\ref{synth-gc-ter5}) 
can be described as having two peaks: the high energy peak (at photon energies above 1~TeV), and the low 
energy peak (at $\mathcal{E} \sim 10$~GeV). In the presented total GC spectra (the \emph{bottom} panel 
of Fig.~\ref{synth-gc-ter5}) only the high energy peak is visible, the low energy peak is dominated by 
the fading magnetospheric component of the cluster $\gamma$-ray emission. The ICS spectral component has 
slightly different shape and also different level depending on the cluster magnetic field values $B_{\rm GC}$ 
used in calculations (see Sect.~\ref{ics-comp}). For low magnetic field 
strength in the cluster ($B_{\rm GC} \lesssim 1$~$\mu$G) a clear dip between the high energy peak (resulting 
from IC scattering of the starlight photon field) and the low energy peak (resulting from IC scattering of 
the CMB) is visible. The dip gradually disappears when the magnetic filed in the cluster increases ($B_{\rm GC} 
> 1$~$\mu$G), leading to single peaked ICS spectrum for the case where $B_{\rm GC} = 10$ and 30~$\mu$G. 
When we increase the $B_{\rm GC}$ value in the cluster, we see that the level of the ICS emission of 
the cluster increases. For the magnetic field values around 3~$\mu$G the emission reaches maximum. Further 
increase in $B_{\rm GC}$ results in the decrease of the emission level of the ICS component. 
This effect of the \emph{increase-saturation-decrease} of the emission component resulting from 
the up-scattering of the stellar photon field and the CMB on the relativistic electrons diffusing through 
the globular cluster environment can be explained by the synchrotron loses suffered by the
electrons in the high magnetic fields. 
For low values of the magnetic field $B_{\rm GC}$ in the cluster,
diffusion of electrons occurs rather fast ($t_{\rm diff} \simeq 3 \times 10^{12}$~s for particles
with energies $E_{\rm el} = 0.1$~TeV, the cluster magnetic field $B_{\rm GC} = 1$~$\mu$G and the cluster 
half mass radius $R_{\rm h} = 1$~pc) and their IC cooling is less efficient. For strong magnetic 
fields particle diffusion is much slower ($t_{\rm diff} \simeq 9 \times 10^{13}$~s for $B_{\rm GC} = 
30$~$\mu$G). However, as the $B_{\rm GC}$ increases, the synchrotron loses of the particles become larger 
leading to the saturation and decrease in production of the ICS photons.

In Fig.~\ref{synth-gc-ter5} the results of simulation for Ter~5 are also contrasted with the available observational 
data in the MeV to GeV \citep[Fermi/LAT data;][]{abdo10} and TeV \citep[H.E.S.S. data;][]{gchess} energy domain. 
With our simulations we obtain the magnetospheric $\gamma$-ray spectra that may reproduce or underestimate 
the cluster emission observed at energies below 10~GeV (see the results presented in the \emph{bottom} panel 
of Fig.~\ref{synth-gc-ter5}). However, we are unable to reproduce the exact spectral shape observed at TeV 
energies.

The $\gamma$-ray flux calculated in this paper should be considered as a guaranteed lower limit provided 
that the model for high energy processes occurring in the inner pulsar magnetosphere is correct. We conclude 
that TeV $\gamma$-ray observations of GCs  can provide independent constraints on the pulsar models.

\subsection{The case of 47 Tucanae}
\label{47tuc-sim}

The 47 Tucanae globular cluster harbours 23 millisecond pulsars detected so far through radio searches 
\citep[see e.g.,][]{camilo05,manchester91}. Similarly to Ter~5, 47~Tuc has been detected in the MeV-GeV 
energy domain with the Fermi/LAT instrument \citep{abdo10}. Its $\gamma$-ray spectrum can be approximated with the
exponentially cut-off power law with a spectral index $\Gamma \simeq 1.4$ and a cut-off energy $\mathcal{E}_{\rm c}
\simeq 2.2$~GeV. The cluster luminosity in this energy range is $L_{\gamma} \simeq 4.8 \times 10^{34}$~erg s$^{-1}$
for the cluster distance $d$ of 4.0~kpc. No TeV emission has been reported so far from 47~Tuc. However, the upper 
limit to the TeV emission of the cluster was estimated \citep{47tuchess}.

The 47 Tuc cluster visual luminosity is $L_{\rm VIS} \simeq 7.5 \times 10^{5} L_{\odot}$ \citep{harris96}. The core 
radius $R_{\rm c}$, the half-mass radius $R_{\rm h}$ and the tidal radius $R_{\rm t}$ were taken from \citet{harris96}.
The ambient photon fields up scattered via inverse Compton process are the cosmic microwave background 
($U_{\rm CMB} = 0.25$ eV cm$^{-3}$) and the photons from the stellar population in 
the cluster (the energy density was estimated with eqs.~4 and 5 of BS07).

\begin{figure}
	\begin{center}
	\begin{tabular}{c}
	\includegraphics[scale=0.44]{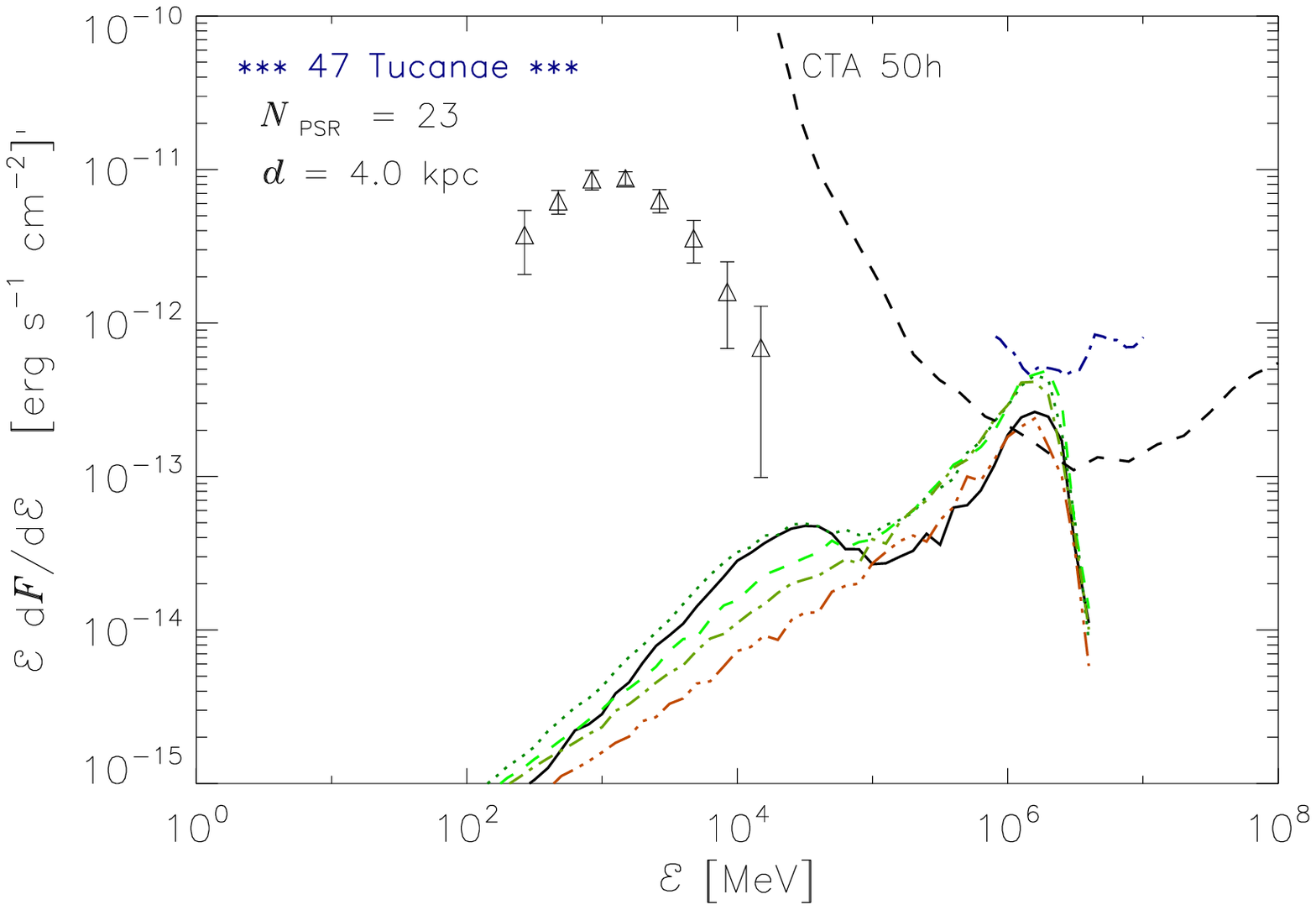}\\
	\includegraphics[scale=0.44]{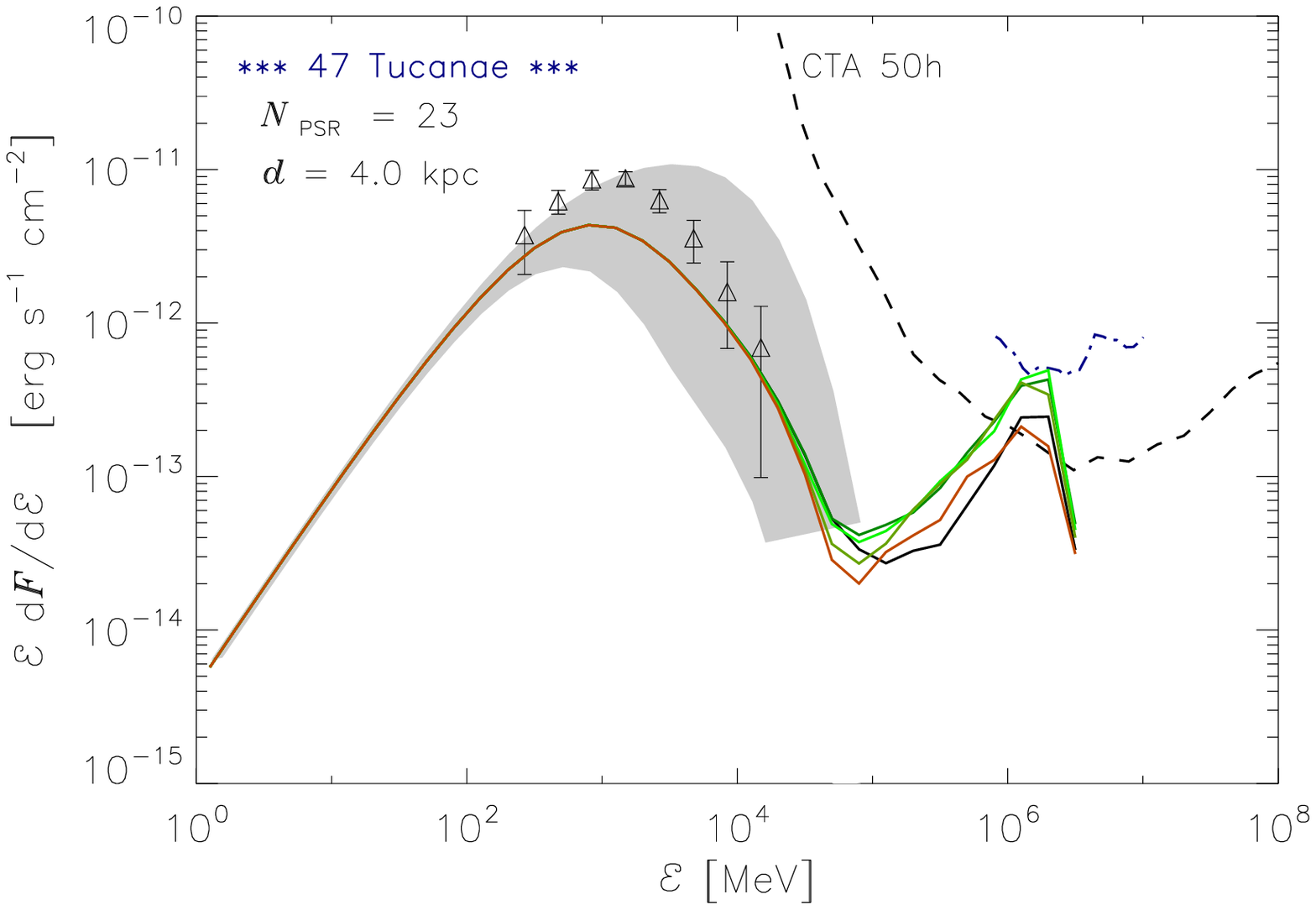}\\
	\end{tabular}
	
	\caption{\small Same as Fig.~\ref{synth-gc-ter5} but for the globular cluster 47 
	Tucanae. Blue dot-dashed line represents the upper limit on TeV emission of 47 Tuc 
	as obtained from H.E.S.S. telescope observations \citep{47tuchess}. Black dashed line 
	shows the CTA differential sensitivity curve with 50~h of integration
	\citep{cta11}.}
	\label{synth-gc-47tuc}
	\end{center}
\end{figure}

All of the detected 23 MSPs in 47~Tuc have spin periods smaller than 11~ms, so $N_{\rm PSR} = 23$ in the simulations.
The results are presented in Fig.~\ref{synth-gc-47tuc}. Similarly as for Ter~5, only high energy peak of the ICS component
is visible in the cluster total spectrum (see the \emph{bottom} panel of Fig.~\ref{synth-gc-47tuc}). The low energy 
peak, which is visible in the ICS spectra presented in the \emph{top} panel of Fig.~\ref{synth-gc-47tuc}, 
is dominated by the magnetospheric component. The \emph{increase-saturation-decrease} behaviour of the ICS emission 
component is visible, though the level difference between the ICS spectrum obtained for the lowest value of the
magnetic field $B_{\rm GC} = 0.3$~$\mu$G and the spectra obtained for higher $B_{\rm GC}$ is smaller than 
in the case of Ter~5.

The simulated $\gamma$-ray spectra are compared with the available observational data from the Fermi/LAT
instrument \citep{abdo10}. Our results rather nicely reproduce the cluster spectrum observed for energies
$\lesssim 10$~GeV. As the cluster has not been detected in the TeV domain, we plot the H.E.S.S. upper limit
\citep{47tuchess}. The modelled ICS spectral components do not violate the H.E.S.S. upper limit. The modelled
TeV emission of 47 Tuc falls right below the presented upper limit. The differential sensitivity
curve for the Cherenkov Telescope Array \citep[CTA;][]{cta11}, the next generation system of Cherenkov 
telescopes, is also displayed in Fig.~\ref{synth-gc-47tuc}. The curve shows the CTA sensitivity obtained 
after 50~h of integration. Our simulations show that the cluster should be detectable with the CTA for all 
of the studied $B_{\rm GC}$. However, the ICS emission is going to be more pronounced for the moderate values 
of the magnetic field within the cluster ($B_{\rm GC}$ between 1~$\mu$G and 10~$\mu$G).

\section{Conclusions}
\label{conlude}

The $\gamma$-ray spectra (energies above $\sim100$~MeV up to $\sim 10$~TeV) for synthetic globular clusters were
calculated. The magnetospheric component (photon energies between $\sim100$~MeV and $\sim 30$~GeV) of the cluster
emission results from the cumulative $\gamma$-ray emission of millisecond pulsars residing within
the cluster core. The ICS component (photon energies above $\sim 30$~GeV) is produced by the ambient photons 
(the stellar photon field originating from stars within the cluster, and the cosmic microwave background) 
up-scattered via inverse Compton process on relativistic electrons injected into the cluster environment 
from the synthetic MSP magnetospheres. The detailed simulations were performed for two globular clusters, 
Terzan~5 and 47~Tucanae, for which observational data are available in the energy range between MeV and TeV. 
Our calculations are able to reproduce, in both cases, the spectrum observed with the Fermi/LAT instrument 
for energies $\lesssim 10$~GeV \citep{abdo09, abdo10}. 
However, for Ter~5 where TeV observations are available, our simulations cannot account for the observed 
spectral shape in this energy domain.

On the other hand, it is possible that colliding pulsar winds, as proposed by BS07, can accelerate leptons 
to relativistic energies. Note that other compact objects within the GCs, such as rotating white dwarfs 
\citep{bednarek12} or accreting neutron stars, could be responsible for acceleration of electrons and their 
injection into the volume of GCs.
The energy spectra of these shock-accelerated electrons 
are power-law in nature. Inclusion of this additional source of relativistic leptons - of different
spectral character than the ones produced in MSP magnetospheres - into the calculations of ICS component
could improve the match between the synthetic TeV spectrum calculated for Ter 5 and the H.E.S.S. 
observations \citep{gchess}. Thus, the calculated synthetic spectra of globular clusters at TeV
energies should be treated as a lower limit to GCs emission in this energy range. This lower limit
can be further tested by the future Cherenkov telescopes like CTA \citep{cta11}.


\section*{Acknowledgments}
This research was partially supported by NCN grants: 2011/01/N/ST9/00485,
2011/01/B/ST9/00411, and DEC-2011/02/A/ST9/00256.

\bibliographystyle{mn2e}
\bibliography{globular_clusters}


\label{lastpage}

\end{document}